\newcommand{\rev}[1]{#1}
\documentclass[acmsmall,screen]{acmart}

\AtBeginDocument{%
  \providecommand\BibTeX{{%
    Bib\TeX}}}

\setcopyright{cc}
\setcctype{by}
\acmJournal{PACMPL}
\acmYear{2026} 
\acmVolume{10} 
\acmNumber{OOPSLA1} 
\acmArticle{123}
\acmMonth{4} 
\acmDOI{10.1145/3798231}

\usepackage{listings}
\usepackage{xcolor}
\lstdefinestyle{code}{%
  backgroundcolor=\color{black!3},
  basicstyle=\ttfamily\small,
  commentstyle=\color{teal!70},
  keywordstyle=\color{blue!70!black}\bfseries,
  stringstyle=\color{orange!70!black},
  numberstyle=\tiny\color{gray},
  numbers=left,
  numbersep=5pt,
  xleftmargin=12pt,
  xrightmargin=4pt,
  frame=single,
  rulecolor=\color{black!20},
  breaklines=true,
  tabsize=2,
  showstringspaces=false,
  columns=fullflexible
}
\usepackage[caption=false]{subfig}
\usepackage[nameinlink,capitalize]{cleveref}
\usepackage{tabularx}
\usepackage[most]{tcolorbox}
\newtcolorbox{takeawaybox}{colback=gray!10, colframe=gray!40, arc=1.5mm, boxrule=0.5pt, left=4pt, right=4pt, top=1pt, bottom=1pt, before skip=4pt, after skip=4pt}

\usepackage{mathtools}
\usepackage{pifont}
\usepackage{multirow}
\usepackage{siunitx}

\def\BibTeX{{\rm B\kern-.05em{\sc i\kern-.025em b}\kern-.08em
    T\kern-.1667em\lower.7ex\hbox{E}\kern-.125emX}}

\def\pname{cuFuzz}
\def\pnameNoDCov{cuFuzz-noDeviceCoverage}
\def\pnameNoSan{cuFuzz-noSanitizer}
\def\pnamePersistent{cuFuzz-persistent}
\def\pnameAFL{AFL++}

\crefformat{figure}{#2Figure~#1#3}
\crefformat{table}{#2Table~#1#3}
\crefformat{listing}{#2Listing~#1#3}
\crefformat{algorithm}{#2Algorithm~#1#3}
\crefformat{equation}{#2Equation~#1#3}
\crefformat{footnote}{#2\footnotemark[#1]#3}

\newcommand{\cmark}{\ding{51}}
\newcommand{\xmark}{\ding{55}}
\newcommand{\cone}{{\large\ding{182}}}
\newcommand{\ctwo}{{\large\ding{183}}}
\newcommand{\cthree}{{\large\ding{184}}}
\newcommand{\cfour}{{\large\ding{185}}}
\newcommand{\cfive}{{\large\ding{186}}}
\newcommand{\csix}{{\large\ding{187}}}
\newcommand{\cseven}{{\large\ding{188}}}
\newcommand{\ceight}{{\large\ding{189}}}

\newcommand{\wone}{{\large\ding{192}}}
\newcommand{\wtwo}{{\large\ding{193}}}
\newcommand{\wthree}{{\large\ding{194}}}

\begin{document}

\title{Hunting CUDA Bugs at Scale with \pname{}}

\author{Mohamed Tarek Ibn Ziad}
\email{mtarek@nvidia.com}
\orcid{0000-0001-6971-6996}
\affiliation{%
  \institution{NVIDIA}
  \city{Westford}
  \state{MA}
  \country{USA}
}
\author{Christos Kozyrakis}
\email{ckozyrakis@nvidia.com}
\orcid{0000-0002-3154-7530}
\affiliation{%
  \institution{NVIDIA}
  \city{Santa Clara}
  \state{CA}
  \country{USA}
}
\affiliation{%
  \institution{Stanford University}
  \city{Stanford}
  \state{CA}
  \country{USA}
}

\begin{abstract}
    
    GPUs play an increasingly important role in modern software. However, the heterogeneous host-device execution model and expanding software stacks make GPU programs prone to memory-safety and concurrency bugs that evade static analysis. While fuzz-testing, combined with dynamic error checking tools, offers a plausible solution, it remains underutilized for GPUs. In this work, we identify three main obstacles limiting prior GPU fuzzing efforts: (1) kernel-level fuzzing leading to false positives, (2) lack of device-side coverage-guided feedback, and (3) incompatibility between coverage and sanitization tools. We present \pname{}, the first CUDA-oriented fuzzer that makes GPU fuzzing practical by addressing these obstacles. 
    
    \pname{} uses whole program fuzzing to avoid false positives from independently fuzzing device-side kernels. It leverages NVBit to instrument device-side instructions and merges the resultant coverage with compiler-based host coverage. Finally, \pname{} decouples sanitization from coverage collection by executing host- and device-side sanitizers in separate processes. \pname{} uncovers 43 previously unknown bugs (19 in commercial libraries) across 14 CUDA programs, including illegal memory accesses, uninitialized reads, and data races. \pname{} achieves significantly more discovered edges and unique inputs compared to baseline approaches, especially on closed-source targets. Moreover, we quantify the execution time overheads of the different \pname{} components and add persistent-mode support to improve the overall fuzzing throughput. Our results demonstrate that \pname{} is an effective and deployable addition to the GPU testing toolbox. \pname{} is publicly available at \url{https://github.com/NVlabs/cuFuzz/}.
    
    \end{abstract}

\settopmatter{printfolios=true}
\settopmatter{printacmref=true}



\received{10 October 2025}
\received[revised]{03 February 2026}
\received[accepted]{17 February 2026}

\maketitle

\section{Introduction} 

Graphics Processing Units (GPUs) play a critical role in modern computing. They serve as core components across diverse fields, from medical image processing~\cite{Litjens2017:DeepLearning} to large language models~\cite{Shoeybi2019:MegatronLM} and quantum computing~\cite{Suzuki2020:Qulacs}. As GPUs become larger and more complex, programming them becomes increasingly challenging and error-prone. Recent studies show that memory safety issues, plaguing CPUs for decades~\cite{Szekeres2013:EternalWar}, also affect GPUs~\cite{Park2021:MindControl,Guo2024:GPU-Exploitation,Roels2025:CUDAattacks,Roh2026:GHOST-ATTACK}. While there are excellent tools to find memory safety bugs in GPU programs either statically~\cite{Betts2012:GPUVerify,Coverity} or dynamically~\cite{Nvidia:ComputeSanitizer,Aditya2021:IGUARD,Ziad2023:cuCatch}, these tools have their own limitations. Static tools suffer from high false positive rates whereas dynamic tools have limited scope—they only detect errors triggered by specific program inputs. When error-triggering inputs are not exercised, dynamic tools miss bugs entirely. A natural solution to this problem is fuzz-testing.

Fuzz-testing (or simply fuzzing) is a well-established CPU testing technique spanning decades~\cite{Miller1990:UNIXFuzz}. It automatically generates random inputs to trigger unexpected program behaviors like crashes and assertions. Fuzzing proves highly effective for uncovering bugs in C/C++ programs~\cite{Google:OSSFuzz}, yet remains largely unexplored for GPUs. Traditional wisdom viewed GPUs as simple accelerators executing well-structured code requiring minimal testing. This perspective no longer holds. Modern GPUs support numerous features, primitives, and operations, making GPU workloads increasingly error-prone. However, naively applying fuzzing to GPUs faces three key challenges: 

\textbf{\wone{} Fuzzing granularity}. GPU programs consist of host-side functions (running on the CPU) and device-side kernels (running on the GPU). Fuzzing can target whole programs or individual kernels. Prior work explored fuzzing individual device kernels by permuting input arguments to expose runtime bugs~\cite{Li2022:CVFuzz} or increase coverage~\cite{Peng2020:OpenCLFuzz}. However, kernel-level fuzzing generates false alarms by creating invalid input combinations that would never occur in complete program execution. For example, a host-side check can invalidate certain device-side input combinations, which kernel-level fuzzing cannot capture.  

\textbf{\wtwo{} Lack of device-side coverage}. Coverage information is essential for guiding the fuzzer. State-of-the-art CPU fuzzers (like AFL++~\cite{Fioraldi2020:AFLplusplus}) use compiler-level instrumentation to collect edge coverage from control flow transitions in the target program. For GPUs, host-side coverage alone is insufficient as it misses critical device-side kernel behaviors. Further, obtaining device-side coverage presents two challenges. First, host and device code use different compilation toolchains (gcc/clang for host, nvcc/ptxas for device). Second, GPUs heavily rely on closed-source libraries~\cite{Nvidia:CudaXLibraries} that cannot be recompiled for coverage instrumentation. A unified approach collecting coverage from both host and device sides is essential for effective GPU fuzzing.

\textbf{\wthree{} Tool incompatibility}. Effective fuzzing requires sanitizers for runtime error detection. While Google's AddressSanitizer (ASan)~\cite{Serebryany2012:ASAN} dominates CPU testing, no ASan equivalent exists for CUDA due to closed-source GPU libraries and compilers. NVIDIA's Compute Sanitizer~\cite{Nvidia:ComputeSanitizer} serves as the alternative, using dynamic binary instrumentation to detect memory safety violations, data races, and uninitialized memory usage. Unfortunately, these sanitizers are neither compatible with each other nor compatible with the coverage collection tools due to shared GPU runtime API dependencies. Hence, coverage collection and sanitization cannot coexist in the same process. This incompatibility severely limits fuzzing effectiveness for GPU code. For example, prior GPU fuzzing work omits sanitizers entirely~\cite{Peng2020:OpenCLFuzz,Li2022:CVFuzz}.

\begin{figure}[!t] 
    \centering
    \includegraphics[width=0.95\textwidth]{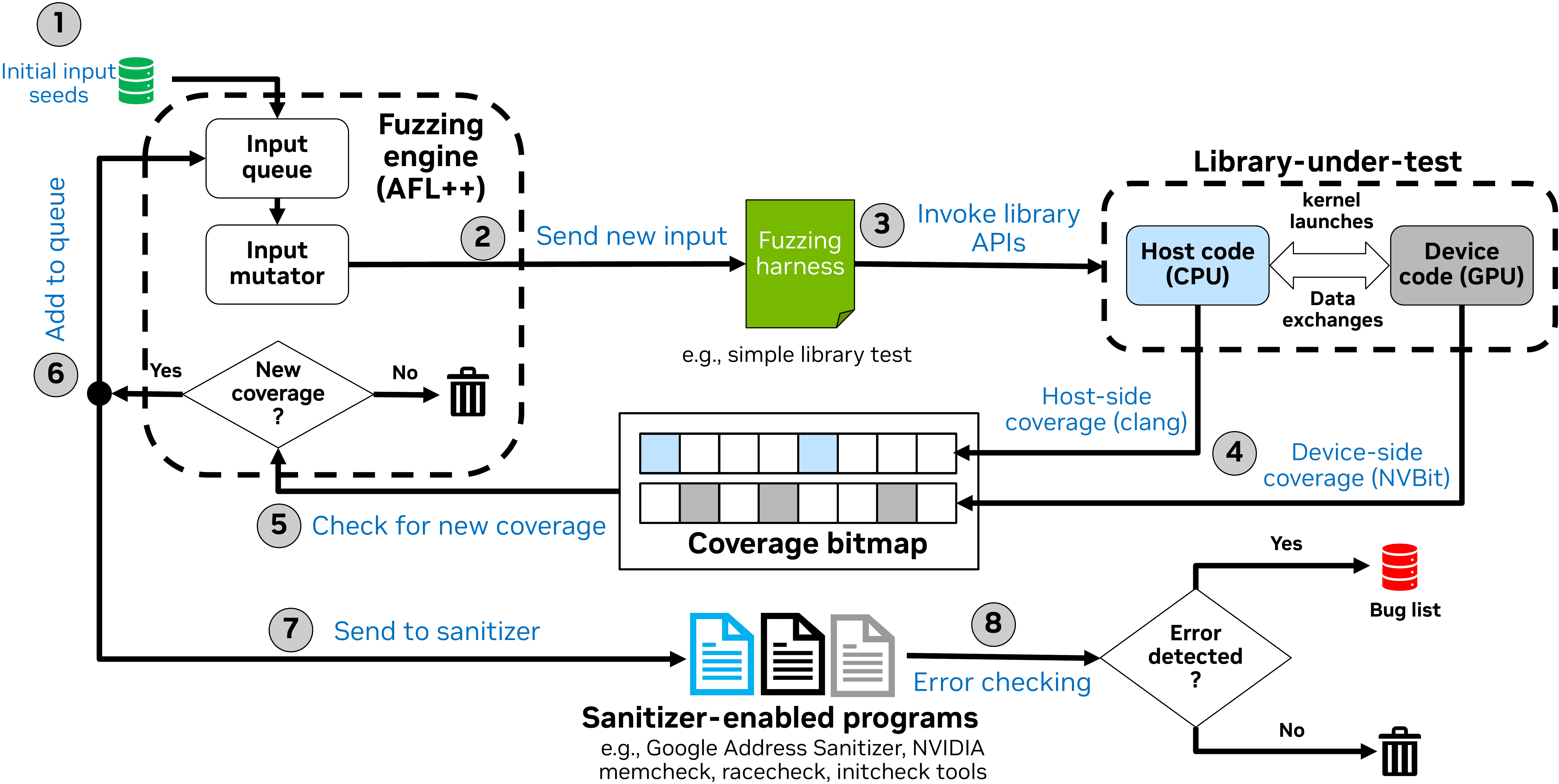}
    \caption{\pname{} high-level overview \rev{showing the integration of AFL++ with NVBit for device-side coverage and various sanitizers (e.g., Compute Sanitizer's memcheck, racecheck, initcheck) for GPU error detection}.} \label{fig:cufuzz-overview} 
    \Description{\pname{} high-level overview.}
    \vspace{-0.15in}
\end{figure}

\textbf{Our work}. We propose \pname{}, the first end-to-end CUDA-oriented fuzzer (\cref{fig:cufuzz-overview}). \pname{} addresses the aforementioned challenges as follows. For challenge~\wone{}, \pname{} operates at the whole program level, avoiding false alarms from kernel-level input permutation while preserving inter-kernel and host-device dependencies. For challenge~\wtwo{}, \pname{} leverages NVBit~\cite{Oreste2019:NVBit}, a dynamic binary instrumentation tool, to collect device-side coverage at runtime. \pname{} merges this with host-side coverage to guide comprehensive fuzzing. For challenge~\wthree{}, \pname{} decouples sanitization from coverage collection by executing sanitizers in separate processes within the fuzzing loop. 

\textbf{Implementation}. We implement \pname{} using AFL++~\cite{Fioraldi2020:AFLplusplus} and NVIDIA's NVBit~\cite{Oreste2019:NVBit}. We evaluate \pname{} on 14 diverse CUDA programs from HeCBench~\cite{Jin2023:HeCBenchPaper,Jin2023:HeCBenchRepo} and CUDALibrarySamples~\cite{Nvidia2025:CUDALibrarySamples}. Our workloads span open-source programs and closed-source libraries (nvTIFF, nvJPEG, nvJPEG2000, cuDNN, cuBLAS). We perform a thorough evaluation of \pname{} to assess its bug finding capabilities, edge and input coverage, performance, sanitization, and efficiency.

\textbf{Results}. From a bug-finding perspective, at the time of writing, \pname{} discovered 43 previously unknown bugs, including 19 in production libraries. These bugs range from memory safety violations (host-side heap overflows, device-side illegal memory accesses) to uninitialized memory reads and data races. We reported all bugs to the corresponding developers who acknowledged them. So far, 40 of 43 bugs have been fixed by the libraries' maintainers in latest releases.

To better understand \pname{}'s individual component contributions, we conduct isolated fuzzing campaigns with only a subset of~\pname{}'s components enabled. We observe that \pname{} achieves significantly more discovered edges and unique inputs compared to vanilla AFL++, especially on closed-source libraries. Additionally, \pname{}'s ability to run various incompatible sanitizers allows it to discover more bugs than AFL++. 

Furthermore, we quantify \pname{}'s execution overheads by measuring the throughput of its individual components. We observe that device-side coverage collection (NVBit) reduces throughput by 67\% whereas host-side coverage collection and sanitization have negligible impact on throughput. On the other hand, device-side sanitization has a different impact on throughput depending on the enabled tool with memcheck, racecheck, and initcheck reducing the throughput by 39\%, 66\%, and 32\%, respectively. Based on this analysis, we share the following recommendations. First, since device-side coverage most benefits closed-source libraries, we recommend disabling it when whole-program recompilation is feasible to avoid the NVBit overheads. Second, to reduce the execution overheads of device-side sanitization, we leverage the insights of a recent work~\cite{Kong2025:SAND} to only run the sanitizers on a subset of inputs (e.g., inputs triggering unique execution paths). 

Our experiments show that CUDA fuzzing is throughput limited. To alleviate this problem, we add support for persistent mode fuzzing in \pname{}. Persistent mode improved performance (i.e., achieved higher coverage in less time) for 50\% of the workloads, discovered the second-highest number of bugs (35 out of 43), and found 16 of them faster than all other \pname{} configurations. \rev{Finally, on the nine open-source benchmarks where kernel-level fuzzing is applicable, \pname{}'s whole-program approach demonstrates clear advantages: kernel-level fuzzing found only 6 of 14 device-side bugs (missing 8 due to violated host-enforced invariants) while producing 16 false positives, whereas \pname{} found all 14 device-side bugs plus~10 additional host-side bugs with zero false positives.} These results demonstrate \pname{}'s value as a practical GPU testing tool. 

\section{Background} \label{sec:background}

This section provides background on GPU architecture and fuzzing. While our work applies to any GPU and fuzzing engine, we focus on NVIDIA GPUs and the American Fuzzy Lop (AFL) due to their widespread adoption.

\subsection{GPU Architecture} \label{subsec:bck-gpu-architecture}
NVIDIA GPUs use the CUDA programming model, which has the following characteristics. 

\subsubsection{Programming Model} \label{subsec:bck-programming-model}
A CUDA program consists of host-side code (running on the CPU) and device-side code (running on the GPU). Host-side code comprises regular C++-like functions, while device-side code consists of special GPU functions called kernels. GPU kernels spawn thousands of threads---the smallest execution units on the GPU. Each thread has its own thread ID and execution context. Threads are organized into warps of 32 threads that execute in Single Instruction, Multiple Threads (SIMT) style on streaming multiprocessors (SMs), the primary programmable units on NVIDIA GPUs. Each SM provides a large register file (e.g., \SI{256}{KB}) and an L1 cache (e.g., \SI{96}{KB}) configurable as on-chip shared memory, accessible only to threads on that SM. All SMs connect to a unified L2 cache, which links to multiple memory controllers for high-bandwidth DRAM access. 

\subsubsection{Memory Spaces} \label{subsec:bck-memory-spaces}
GPUs have multiple memory spaces: \texttt{local} memory (available only to individual threads), \texttt{shared} memory (shared between threads on the same SM), and \texttt{global} memory (accessible to all GPU threads). Additional memory spaces include constant memory and texture memory, both limited to read-only access and used for specialized data access patterns and graphics rendering, respectively. 

\subsection{Fuzzing and Sanitization} \label{subsec:bck-fuzzing}
Fuzzing is a widely used software testing method. It generates inputs (based on pre-defined rules or random strategies) and feeds them to the target program. The program executes with these inputs while being monitored for crashes or abnormal outputs. This process repeats with newly generated inputs for extended periods (hours or days). Inputs causing crashes are saved for analysis to identify vulnerability root causes. Fuzzing can be divided into three main categories: \textit{black-box fuzzing} (no knowledge of program internals, monitoring only outputs), \textit{white-box fuzzing} (complete program knowledge, using techniques like taint analysis or symbolic execution), and \textit{grey-box fuzzing} (partial program knowledge, using code coverage to guide input generation).

\subsubsection{AFL++} \label{subsec:bck-aflpp}
AFL++ is the state-of-the-art coverage-guided grey-box fuzzer~\cite{Fioraldi2020:AFLplusplus}. This community-driven fuzzer has been extensively used in academia and industry. Starting with a fuzzing harness and initial seed inputs, AFL++ uses multiple mutation strategies (byte flips, arithmetic operations, input merging, dictionary-based mutations) to mutate program inputs. AFL++ uses compiler instrumentation to track branch (edge) coverage between basic blocks. At runtime, it stores executed edges for each input in a coverage bitmap. This bitmap is compared against a reference bitmap storing all previously seen edges to identify inputs triggering new code paths. Inputs that lead to new edges are saved in the fuzzing queue for further mutation. While AFL++ has a QEMU mode for closed-source CPU library fuzzing, it lacks CUDA program support, making host-side code instrumentation the only feasible option when fuzzing GPU programs. 

\subsubsection{Sanitizers} \label{subsec:bck-sanitizers}
Fuzzing typically involves running error checking tools (also known as sanitizers) to detect errors that do not cause program crashes. State-of-the-art sanitizers include Google's AddressSanitizer (ASan~\cite{Serebryany2012:ASAN}), a compiler-based tool checking for memory safety violations like buffer overruns and use-after-free in C/C++ code, and NVIDIA's Compute Sanitizer~\cite{Nvidia:ComputeSanitizer}, a dynamic binary instrumentation tool detecting various device-side errors. The tools supported by Compute Sanitizer include memcheck for catching memory safety errors, initcheck for detecting uninitialized memory accesses, and racecheck for detecting data races, mainly in shared memory space. 
\section{Motivation} \label{sec:motivation}

This section motivates the need for GPU-aware fuzzing by highlighting the key challenges in GPU-based testing. We illustrate these challenges using a simple CUDA example (\cref{fig:vector-add}) containing four device-side kernels (\texttt{clamp}, \texttt{scaleInPlace}, \texttt{vectorAdd}, \texttt{checksumKernel}) and one host-side function (\texttt{validateInputs}). The \texttt{main} function accepts two data buffers and their sizes as inputs. It validates inputs using \texttt{validateInputs}, allocates device memory, launches the four kernels, and copies the final checksum back to the host. This example demonstrates GPU program complexity and why traditional fuzzing approaches prove insufficient.

\begin{figure}[t]
\begin{minipage}[t]{0.49\textwidth}
\begin{lstlisting}[language=C++, caption=CUDA motivating example., label=fig:vector-add, basicstyle=\tiny, escapeinside={(*@}{@*)}]
// Host-side validation function
bool validateInputs(int n, float* a, float* b) {
    if (n <= 0 || n > MAX_VECTOR_SIZE) return false;(*@\label{line:motiv-line3}@*)
    if (!a || !b) return false;(*@\label{line:motiv-line4}@*)
    return true;
}

// Device kernel with off-by-one bounds bug
__global__ void vectorAdd(float* a, float* b, float* c, int n) {
    int idx = blockIdx.x * blockDim.x + threadIdx.x;
    if (idx <= n) {  // BUG: off-by-one allows idx == n(*@\label{line:motiv-line11}@*)
        c[idx] = a[idx] + b[idx];(*@\label{line:motiv-line12}@*)
        // Complex device-side logic with edge cases
        if (c[idx] > 1e6) {(*@\label{line:motiv-line14}@*)
            // Edge case: large values trigger special processing
            c[idx] = processLargeValue(c[idx]);(*@\label{line:motiv-line16}@*)
        }
    }
}

__global__ void clamp(float* x, int n, float lo, float hi) {
    int idx = blockIdx.x * blockDim.x + threadIdx.x;
    if (idx < n) {
        float v = x[idx];
        if (v < lo) v = lo;
        if (v > hi) v = hi;
        x[idx] = v;
    }
}

__global__ void scaleInPlace(float* x, int n, float factor) {
    int idx = blockIdx.x * blockDim.x + threadIdx.x;
    if (idx < n) {
        x[idx] *= factor;
    }
}

__global__ void checksumKernel(const float* x, int n, 
    unsigned long long* out) {
    int idx = blockIdx.x * blockDim.x + threadIdx.x;
    if (idx < n) {
        // Simple checksum: sum of absolute integer parts
        unsigned long long v = 
                (unsigned long long)fabs((double)x[idx]);
        atomicAdd(out, v);
    }
}
\end{lstlisting}
\end{minipage}\hfill
\begin{minipage}[t]{0.49\textwidth}
\begin{lstlisting}[language=C++, firstnumber=last, basicstyle=\tiny, framexrightmargin=-5pt]
// Program entry point
int main(int argc, char** argv) {
    // Read program inputs (n, and host buffers a, b). Details elided.
    int n = /* read from input */ 0;
    float *a = /* read host buffer */ nullptr;
    float *b = /* read host buffer */ nullptr;
    float *c = (float*)malloc((size_t)n * sizeof(float));

    // Host-side validation
    if (!validateInputs(n, a, b)) {
        return 1;
    }

    // Allocate device memory
    float *d_a, *d_b, *d_c;
    cudaMalloc(&d_a, (size_t)n * sizeof(float));
    cudaMalloc(&d_b, (size_t)n * sizeof(float));
    cudaMalloc(&d_c, (size_t)n * sizeof(float));
    
    unsigned long long *d_checksum; 
    cudaMalloc(&d_checksum, sizeof(unsigned long long));
    cudaMemset(d_checksum, 0, sizeof(unsigned long long));

    // Copy data to device
    cudaMemcpy(d_a, a, (size_t)n * sizeof(float), cudaMemcpyHostToDevice);
    cudaMemcpy(d_b, b, (size_t)n * sizeof(float), cudaMemcpyHostToDevice);
    
    // Launch a sequence of kernels
    int blockSize = 256;
    int gridSize = (n + blockSize - 1) / blockSize;
    // Preprocess inputs on device 
    clamp<<<gridSize, blockSize>>>(d_a, n, -1e3f, 1e3f);
    scaleInPlace<<<gridSize, blockSize>>>(d_b, n, 0.5f);
    // Core computation 
    vectorAdd<<<gridSize, blockSize>>>(d_a, d_b, d_c, n);
    // Postprocess/validate result 
    checksumKernel<<<gridSize, blockSize>>>(d_c, n, d_checksum);
    // Copy result back
    cudaMemcpy(c, d_c, (size_t)n * sizeof(float), cudaMemcpyDeviceToHost);

    // Cleanup
    cudaFree(d_a); cudaFree(d_b);
    cudaFree(d_c); cudaFree(d_checksum);
    free(c);
    return 0;
}
\end{lstlisting}
\end{minipage}
\Description{CUDA motivating example.}
\end{figure}

\subsection{Kernel-Level versus Whole-Program Fuzzing} \label{subsec:challenge1-whole-program}
Prior work explored fuzzing individual device-side kernels by permuting input arguments. For example, the \texttt{vectorAdd} kernel can be fuzzed by independently permuting buffers \texttt{a}, \texttt{b}, and argument~\texttt{n}. However, this approach generates false alarms by creating invalid input combinations that never occur in complete program execution. This can be demonstrated by two examples: (1)~Providing NULL pointers causes illegal memory access in \hyperref[line:motiv-line12]{line~12}, but host-side validation~(\hyperref[line:motiv-line4]{line~4}) already prevents this. (2)~Providing negative \texttt{n} values causes out-of-bounds access, but input validation (\hyperref[line:motiv-line3]{line~3}) blocks this scenario. In short, kernel-level fuzzers, which ignore host-side logic, will incorrectly flag issues that never occur in full programs. Conversely, kernel-level fuzzing misses bugs whose root cause lies in host-device interactions, e.g., when unchecked CUDA API failures cause device-side errors that cannot be reproduced in isolation.
\begin{takeawaybox}
\textit{A whole-program fuzzer is essential for uncovering realistic bugs in CUDA programs while avoiding false positives from kernel-level input permutation}.
\end{takeawaybox}

\subsection{Lack of Device-Side Coverage} \label{subsec:challenge2-device-side-coverage}
Coverage information plays a crucial role in guiding fuzzing. While host-side coverage can indicate device-side execution (when host functions launch device kernels), it may not capture all unique device-side executions. Consider the \texttt{vectorAdd} kernel in \cref{fig:vector-add}. Random inputs with valid~\texttt{n},~\texttt{a},~\texttt{b} values exercise most host-side code and will accidentally cover some device-side conditions (e.g., \hyperref[line:motiv-line11]{line~11}). However, \hyperref[line:motiv-line14]{lines~14--17} contain conditional logic, \texttt{processLargeValue()}, which is only executed under specific GPU conditions. Host-side fuzzing alone cannot exercise these GPU-only paths since it is unaware of the condition at \hyperref[line:motiv-line14]{line~14}. 
\begin{takeawaybox}
    \textit{An efficient GPU fuzzer must gather coverage data (such as executed edges) from both host- and device-side code}.
\end{takeawaybox}

\subsection{Incompatibility Between Different Fuzzing Components} \label{subsec:challenge3-incompatibility}
Sanitizers are key fuzzing components that uncover issues missed during baseline execution. The \texttt{vectorAdd} kernel in \cref{fig:vector-add} contains a bug in \hyperref[line:motiv-line11]{line~11}: \texttt{if (idx $\le$ n)} should be \texttt{if (idx $<$ n)}. This creates an off-by-one error in \hyperref[line:motiv-line12]{line~12}. The overflow is too small to trigger a GPU runtime segmentation fault, so it goes unnoticed even with Google's AddressSanitizer, which only validates host-side memory accesses. However, NVIDIA Compute Sanitizer's memcheck tool easily captures this overflow by tracking exact buffer bounds and validating all device memory accesses. Unfortunately, enabling device-side sanitizers is challenging since they depend on the same runtime APIs as other GPU tools (NVBit, cuda-gdb). Only one tool can be enabled simultaneously. 
\begin{takeawaybox}
    \textit{Running device-side sanitizers during fuzzing is strongly recommended to maximize benefits, but they conflict with each other and with coverage collection tools}.  
\end{takeawaybox}   

\subsection{Throughput} \label{subsec:challenge4-throughput}
While whole-program fuzzing offers significant benefits, it incurs substantial costs. CUDA runtime initialization creates a major bottleneck, adding overhead to each input trial. Additionally, enabling multiple sanitizers within the fuzzing loop further increases runtime costs, resulting in very low throughput as will be demonstrated in our experiments. 
\begin{takeawaybox}
    \textit{An efficient GPU fuzzer should maximize throughput while strategically enabling sanitizers to catch more errors}. 
\end{takeawaybox}

\section{\pname{} High-Level Design} \label{sec:design}

This section presents \pname{}, the first CUDA-oriented fuzzer. The description here is tool-agnostic as implementation details are provided in \cref{sec:implementation}. 

\subsection{Overview} \label{subsec:overview}

\cref{fig:cufuzz-overview} shows \pname{}'s high-level overview. \pname{} comprises four main components: (a)~a fuzzing harness (the fuzzing entry point), (b)~a fuzzing engine (generates and selects new inputs), (c)~coverage collection (gathers coverage information from the program under test), and (d)~sanitization (performs error detection). While these components exist in any fuzzing framework, \pname{} uniquely positions them to address the challenges discussed in \cref{sec:motivation}. 

\subsection{Whole Program Fuzzing} \label{subsec:soln1-whole-program-fuzzing}
To avoid false positives from permuting individual kernel inputs without considering inter-kernel dependencies or host-side code, \pname{} operates at the whole program level. Our fuzzing harness targets entire CUDA programs rather than single kernels. For CUDA applications, we use the application's main function as the harness. For closed-source CUDA libraries, we use accompanying examples as harnesses (e.g., the nvTIFF-Decode-Encode example from CUDALibrarySamples~\cite{Nvidia2025:CUDALibrarySamples}). 

The fuzzing harness is instrumented according to the fuzzing engine (e.g., AFL++ or LibFuzzer) to update coverage information. This instrumentation occurs at compile time (when source code is available) or execution time (for closed-source libraries). Using whole programs as harnesses enables capturing critical bugs arising from CUDA API interactions or shared kernel state (e.g., races occurring when memory accesses from different kernels touch the same memory location with one of them being a write).

\subsection{Device-Side Coverage Collection} \label{subsec:soln2-device-side-coverage-collection}
For closed-source library fuzzing, the harness only contains the API calls without information about the library internals. For instance, the nvTIFF library example has a single encode API triggering multiple kernel launches: \texttt{compactStrips\_k}, \texttt{exsumMax\_1blk\_k}, \texttt{compressStrips\_k}, and \texttt{batchedCopyLittleEndian\_k}, as revealed by dynamic binary instrumentation tools~\cite{Oreste2019:NVBit}. Consequently, host-side coverage alone cannot adequately guide fuzzing. \pname{} collects device-side coverage at runtime using NVBit~\cite{Oreste2019:NVBit}, a dynamic binary instrumentation tool for GPUs. 

\rev{For host-side coverage, \pname{} relies on standard AFL++ compile-time instrumentation.} However, device-side collection poses challenges due to inherent differences between host and device execution models. Host code's single-threaded nature enables simple coverage data structure updates via write operations. On the device side, thousands of threads may execute the same branch instruction simultaneously, attempting to update the same coverage entry. \pname{} adopts several optimizations for this scenario, discussed in \cref{sec:implementation}. Additionally, collisions may occur between host-side coverage entries (assigned at compile time) and device-side entries (assigned at dynamic load time). \pname{} addresses this problem by assigning disjoint regions in the coverage data structure for host and device code.

\subsection{Decoupling Host- and Device-Side Sanitization} \label{subsec:soln3-decoupling-sanitization}
To address incompatibility issues between tools involved in fuzzing (e.g., NVBit and Compute Sanitizer), \pname{} decouples sanitization from coverage collection. We run separate processes for each desired sanitizer within the fuzzing loop, in addition to the main fuzzing harness collecting coverage information. While this approach increases runtime costs (due to multiple processes), overheads are reduced by selectively running sanitizers on input subsets (e.g., inputs with unique execution paths), as will be shown in~\cref{sec:eval}. 

\subsection{Putting It All Together} \label{subsec:putting-it-all-together}
We now put all components together by examining the complete fuzzing loop. Starting from \cref{fig:cufuzz-overview}'s top left, in step \cone{}, users provide a CUDA library to test, a sample program invoking it, and seed inputs (raw data files or images). Next, \ctwo{} the fuzzing engine adds initial inputs to a queue and begins mutating them using its mutation strategies. \cthree{} Mutated inputs are passed to the fuzzing harness, which invokes library APIs. 

At runtime, \cfour{} \pname{} collects host-side coverage (executed edges) through traditional compile-time instrumentation and device-side coverage through custom dynamic binary instrumentation, storing both in the coverage data structure (coverage bitmap). In step \cfive{}, the fuzzing engine compares the per-input coverage bitmap against a local copy, which tracks all previously explored edges. If the per-input bitmap contains no new entries, the input is discarded. However, if at least one new host- or device-side edge is detected, \csix{} the corresponding input is added to the queue for further mutation. \cseven{} We also send this input to sanitizer-enabled versions of the fuzzing harness according to a pre-configured sanitization strategy. \ceight{} Finally, error-triggering inputs are saved to the crashes folder for inspection while benign inputs are discarded.

\section{Implementation} \label{sec:implementation}
This section describes \pname{}'s implementation details. 

\subsection{Fuzzing Loop} \label{subsec:imp-fuzzing-loop}
While \pname{}'s design from \cref{sec:design} can be built on top of any CPU fuzzer, our proof-of-concept implementation uses AFL++ as the fuzzing engine for input mutation. We also use AFL++'s compiler to build the fuzzing harness and insert necessary edge coverage instrumentation. For simplicity, we focus on mutating file-based inputs. 

For example, when fuzzing a \texttt{medianfilter} program accepting a ``.ppm'' image and iteration count, we fix the iteration count at two and permute only the .ppm file contents. For programs not accepting input files (e.g., \texttt{convolution3D}, which accepts six integers: batch size, input channels, output feature maps, input width, input height, and kernel size), we modify the harness interface to read six integers from a file as a series. This preserves AFL++'s interface without modifications.

\subsection{Coverage Collection} \label{subsec:imp-coverage-collection}
\rev{\pname{} uses a two-pronged approach for coverage collection: standard AFL++ instrumentation for host-side code and a custom NVBit tool for device-side code. Host-side coverage requires no custom implementation as AFL++'s compiler inserts edge-tracking instrumentation during compilation. For device-side coverage,} we built an NVBit tool to collect edge-coverage information from device-side code. NVBit is a dynamic binary instrumentation framework that patches device-side kernels at load time to enable dynamic instrumentation of device-side GPU code~\cite{Oreste2019:NVBit}. When the CUDA context is initialized (typically with the first kernel launch), our NVBit tool captures the AFL++ coverage bitmap address and allocates GPU memory buffers to track per-thread executed edges, similar to AFL++'s host-side coverage tracking. 

At kernel load time, our NVBit tool replaces the first instruction of each basic block on the device-side with a jump instruction pointing to trampoline code. In the trampoline, NVBit saves the thread context, executes our coverage-collection instrumentation function (\texttt{cufuzz\_cov\_edge}), restores the program context, executes the original instruction, and jumps back. At the final CUDA context execution, our NVBit tool copies the device-side bitmap to the host and merges it with the host-side bitmap. Due to the inherent differences between host-side and device-side code, our tool addresses the following three challenges.

\begin{figure}[!h]
\centering
\begin{minipage}[t]{0.49\textwidth} 
\begin{lstlisting}[language=C, caption={AFL++ host-side instrumentation.}, label={lst:afl-cov-edge}, basicstyle=\tiny]
extern unsigned char* __afl_area_ptr;     /* points to shared bitmap */
__thread uint32_t __afl_prev_loc = 0;     /* per-thread edge history */

static inline void afl_cov_edge(uint32_t cur_loc) {

    uint32_t idx = __afl_prev_loc ^ cur_loc;

    /* never-zero increment of 8-bit counter */
    uint8_t v = __afl_area_ptr[idx];
    v++;
    __afl_area_ptr[idx] = v ? v : 1;

    __afl_prev_loc = cur_loc >> 1;
}
\end{lstlisting}
\end{minipage}\hfill
\begin{minipage}[t]{0.49\textwidth}
    
\begin{lstlisting}[language=C++, caption={\pname{} device-side instrumentation.}, label={lst:nvbit-cov-edge}, basicstyle=\tiny, escapeinside={(*@}{@*)}]
extern "C" __device__ __noinline__ void cufuzz_cov_edge(
  int cur_loc, uint64_t __afl_area_ptr, uint64_t prev_loc) {
  const int active_mask = __ballot_sync(__activemask(), 1);
  const int laneid = get_laneid();
  const int first_laneid = __ffs(active_mask) - 1;

  uint64_t tid = threadIdx.x + blockIdx.x * blockDim.x
  + threadIdx.y * blockDim.x * gridDim.x
  + blockIdx.y * blockDim.x * blockDim.y * gridDim.x
  + threadIdx.z * blockDim.x * blockDim.y * gridDim.x * gridDim.y
  + blockIdx.z * blockDim.x * blockDim.y * blockDim.z * gridDim.x * gridDim.y;
  int idx = (MAP_SIZE/2) + ((prev_loc[tid] ^ cur_loc) % (MAP_SIZE/2));(*@\label{line:nvbit-mapsize}@*)
  prev_loc[tid] = cur_loc >> 1;
  /* only the first active thread will perform the atomic update*/
  if (first_laneid == laneid) {
    atomicAdd((unsigned int*)&__afl_area_ptr[idx], 1);(*@\label{line:nvbit-atomic}@*)
  }
}
\end{lstlisting}
\vspace{-0.1in}
\end{minipage}

\caption{\pname{} coverage \rev{instrumentation: (Left)~AFL++ host-side edge tracking via compile-time instrumentation; (Right)~NVBit device-side edge tracking via runtime instrumentation with warp-aware atomic updates.}}
\Description{\pname{} coverage instrumentation: (Left)~AFL++ host-side edge tracking via compile-time instrumentation; (Right)~NVBit device-side edge tracking via runtime instrumentation with warp-aware atomic updates.}
\label{fig:cov-side-by-side}
\vspace{-0.1in}
\end{figure}

\subsubsection{Handling Concurrent Updates}
\cref{lst:afl-cov-edge} shows the host-side coverage collection instrumentation function pseudo-code, which XORs the current edge identifier with the previous edge identifier and updates the coverage data structure. Using the same function on the device side causes significant instability since multiple threads can execute it simultaneously. We address this issue using atomic updates as shown in \hyperref[line:nvbit-atomic]{line~16} of \cref{lst:nvbit-cov-edge}. 

\subsubsection{Handling Bitmap Collisions}
Our current prototype uses a \SI{64}{KB} coverage bitmap where each host- or device-side edge is represented by a single byte. To avoid collisions between host- and device-side bitmap entries, host-side coverage entries use the first \SI{32}{KB} while device-side coverage entries use the second \SI{32}{KB}. We achieve this by starting device-side edges at \texttt{MAP\_SIZE/2}, as shown in \hyperref[line:nvbit-mapsize]{line~12} of \cref{lst:nvbit-cov-edge}. 

\subsubsection{Handling Thread Count}
In AFL++ terminology, the byte representing host-side edges in the coverage bitmap not only indicates whether an edge is executed or not, but also tracks execution frequency per input (enabling the identification of edges executed within loops). This count is bucketed to powers of two to avoid path explosion. An input is considered interesting (saved to the queue) if it explores at least one new bucket for an edge. We use the same approach on the device side by bucketing thread counts (how many threads executed the edge). To account for potentially massive GPU thread numbers, we use larger power-of-two buckets and warp-level (instead of thread-level) counts. As shown in \hyperref[line:nvbit-atomic]{line~16} of \cref{lst:nvbit-cov-edge}, a value of one is used to increment the bitmap entry counter for the entire warp.

\subsection{Sanitization} \label{subsec:imp-sanitization}
Inspired by recent work, SAND~\cite{Kong2025:SAND}, we decouple coverage collection from host- and device-side sanitization in \pname{}'s fuzzing loop, avoiding incompatibility issues between NVBit and Compute Sanitizer. The original SAND implementation, now merged with AFL++, assumes each sanitizer-enabled program is instrumented at compile time with sanitizer instrumentation in addition to AFL++ server instrumentation. Since our main device-side sanitizer, NVIDIA Compute Sanitizer, relies on binary instrumentation, we create a wrapper around the program under test that can be compiled with SAND (to add AFL++ server instrumentation) and used to launch both Compute Sanitizer and the original program. While this approach increases execution time costs (due to multiple processes), overheads are amortized by selectively running sanitizers on input subsets during fuzzing, as will be quantified in~\cref{subsec:cufuzz-sanitization-modes}.

\begin{figure}[!h]
\centering
\begin{minipage}[t]{0.49\textwidth}

\begin{lstlisting}[language=C, basicstyle=\tiny, caption={Harness persistent mode modifications.}, label={lst:persistent-harness}, escapeinside={(*@}{@*)}]
__AFL_FUZZ_INIT();

__global__ void cufuzz_notification_kernel(int signal_id) {
    /* Empty kernel - just for NVBit signaling */
    ;
    /* Signal ID can be used to identify different events */
}

// Fuzzing harness entry point
main() {
    /* program initialization. */
    /* ... */

    #ifdef __AFL_HAVE_MANUAL_CONTROL
    __AFL_INIT();
    #endif
    unsigned char *buf = __AFL_FUZZ_TESTCASE_BUF;  

    while (__AFL_LOOP(1000)) {
        int len = __AFL_FUZZ_TESTCASE_LEN;  
        if (len < 512) continue;  /* check for a useful min. length */

        /* write to output file for sanitization */
        FILE *fp = fopen(fname, "wb");
        if (fp != NULL) {   
            fwrite(buf, 1, len, fp); fclose(fp);
        }
        cufuzz_notification_kernel<<<1, 1>>>(1234); /* iteration start */(*@\label{line:persistent-start}@*)
        /* ... */
        /* regular fuzzing harness code */
        /* ... */
        cufuzz_notification_kernel<<<1, 1>>>(5678); /* iteration end */(*@\label{line:persistent-end}@*)
    }
    return 0;
}
\end{lstlisting}
\end{minipage}\hfill
\begin{minipage}[t]{0.49\textwidth}
    
\begin{lstlisting}[language=C++, basicstyle=\tiny, caption={\pname{} persistent mode support in NVBit.}, label={lst:nvbit-persistent}]
void nvbit_at_cuda_event(CUcontext ctx, int is_exit, nvbit_api_cuda_t cbid, const char *name, void *params, CUresult *pStatus) {
    /* Identify all the possible CUDA launch events */
    if (cbid == API_CUDA_cuLaunch || 
        cbid == API_CUDA_cuLaunchKernel_ptsz ||
        cbid == API_CUDA_cuLaunchGrid || 
        cbid == API_CUDA_cuLaunchGridAsync ||
        cbid == API_CUDA_cuLaunchKernel) {
        cuLaunch_params *p = (cuLaunch_params *)params;
        std::string kernel_name = nvbit_get_func_name(ctx, p->f, 1);
        if (kernel_name == "_Z19cufuzz_notification_kerneli" && afl_persistent && !is_exit) {
            /* Check for a specific notification kernel */
            if (cbid == API_CUDA_cuLaunchKernel_ptsz || 
                cbid == API_CUDA_cuLaunchKernel) {
                cuLaunchKernel_params* p_kernel = (cuLaunchKernel_params*)params;
                if (p_kernel->gridDimX == 1 && 
                    p_kernel->gridDimY == 1 && 
                    p_kernel->gridDimZ == 1 &&
                    p_kernel->blockDimX == 1 && 
                    p_kernel->blockDimY == 1 && 
                    p_kernel->blockDimZ == 1) {
                    uint32_t magic_value = *((uint32_t*)p_kernel->kernelParams[0]);
                    if (magic_value == MAGIC_VALUE_START) { 
                        /* reset device-side coverage bitmap */
                    } else if (magic_value == MAGIC_VALUE_END) { 
                        /* merge device and host coverage bitmaps */
                    }
                }
            }
        }
    }
}
\end{lstlisting}
\vspace{-0.1in}
\end{minipage}  
\caption{\pname{} persistent mode\rev{: (Left)~Harness structure with AFL++ loop processing multiple inputs per process; (Right)~State reset between iterations in NVBit.}}
\Description{\pname{} persistent mode\rev{: (Left)~Harness structure with AFL++ loop processing multiple inputs per process; (Right)~State reset between iterations in NVBit.}}
\label{fig:persistent-side-by-side}
\vspace{-0.1in}
\end{figure}

\subsection{Persistent Mode} \label{subsec:imp-persistent-mode}
AFL++ runs one test case per process by default. This avoids test case interference and improves error attribution. However, this creates large overhead for GPU fuzzing due to CUDA runtime's substantial initialization cost. One potential solution is to use AFL++'s persistent mode, allowing the target program to run in a loop, executing one test case per iteration. The pseudo-code for fuzzing harness changes required for AFL++ persistent mode appears in \cref{lst:persistent-harness}. 

To enable persistent mode for \pname{}, we have to address two challenges. First, we need a method to notify NVBit of loop boundaries so our custom tool can differentiate between test cases and update the device-side coverage bitmap accordingly. Second, we need to pass inputs from the persistent run to the separately running sanitizer processes. 

We address the first challenge by calling an empty kernel \texttt{cufuzz\_notification\_kernel} (\hyperref[line:persistent-start]{lines~28} and \hyperref[line:persistent-end]{32} in \cref{lst:persistent-harness}) with a unique identifier at the beginning and end of the persistent loop. We modify our NVBit coverage tool to intercept these unique kernel invocations, check the input parameter, and take appropriate actions (i.e., reset the device bitmap at loop beginning and merge it with the host bitmap at loop end). We address the second challenge by writing the newly generated buffer \texttt{buf} to a file that is read by our sanitizer-enabled wrapper. We evaluate persistent mode coverage and bug finding benefits in~\cref{sec:eval}. 

\section{Experimental Methodology} \label{sec:methodology}

\subsection{Benchmarks} \label{subsec:methodology-benchmarks}
We evaluate 14 programs from the heterogeneous computing benchmark suite HeCBench \cite{Jin2023:HeCBenchPaper,Jin2023:HeCBenchRepo} and CUDA library samples CUDALibrarySamples~\cite{Nvidia2025:CUDALibrarySamples}.\footnote{For HeCBench, we use commit number: 37507304619620b716977d63829871b658465986. For CUDA libraries, we use the official build versions shown in~\cref{tab:cufuzz-benchmarks}.} \rev{These benchmarks span diverse GPU application domains: \emph{Machine Learning} (\texttt{cuDNN}, \texttt{attention}), \emph{Math} (\texttt{cuBLAS}, \texttt{lud}), \emph{Image Processing} (\texttt{nvTIFF}, \texttt{nvJPEG}, \texttt{nvJPEG2000}, \texttt{boxfilter}, \texttt{medianfilter}, \texttt{recursiveGaussian}, \texttt{seam-carving}), \emph{Data Compression/Encoding} (\texttt{dxtc2}, \texttt{crs}), and \emph{Random Number Generation} (\texttt{urng}).} \cref{tab:cufuzz-benchmarks} summarizes our benchmarks, including fuzzing harness, application or library name, domain, and input format. The table includes static binary size as a proxy for program complexity. 

\begin{table}[!t]
    \centering
    \caption{Summary of our benchmarks.}
	\resizebox{0.9\textwidth}{!}{
    \label{tab:cufuzz-benchmarks}
    \begin{tabular}{lllll}
        \hline
        Fuzzing harness & Target library & Domain & Input format & Static binary size \\
        \hline
		attention & attention & Machine learning & 4 integers & \SI{1.3}{MB} \\
		boxfilter & boxfilter & Image processing & ppm & \SI{1.7}{MB} \\
		crs & crs & Data encoding & 2 integers & \SI{1.6}{MB} \\
		dxtc2 & dxtc2 & Data compression & ppm & \SI{1.5}{MB} \\
		lud & lud & Math & 1 integer & \SI{1.2}{MB} \\
		medianfilter & medianfilter & Image processing & ppm & \SI{1.4}{MB} \\
		recursiveGaussian & recursiveGaussian & Image processing & ppm & \SI{1.5}{MB} \\
		seam-carving & seam-carving & Image processing & jpg & \SI{2.0}{MB} \\
		urng & urng & Random number generation & bmp & \SI{1.2}{MB} \\
		\hline
		nvtiff\_example & libnvtiff-0.4.0.62 & Image processing & tif & \SI{4.0}{MB} \\
		nvjpegDecoder & libnvjpeg-12.4.0.33 & Image processing & jpg & \SI{8.8}{MB} \\
		nvjpeg2k\_example & libnvjpeg2k-0.8.0.38 & Image processing & jp2, j2k & \SI{6.4}{MB} \\
		convolution3D & cuDNN-9.11.0 & Neural networks & 6 integers & \SI{29}{MB} \\
        blas-gemm & cuBLAS-12.9.1.4 & Math & 4 integers & \SI{465}{MB} \\
        \hline
    \end{tabular}
	}
    \Description{Summary of our benchmarks.}
	\vspace{-0.1in}
\end{table}

\subsection{Tools} \label{subsec:methodology-tools}
Our implementation uses AFL++ 4.31c, NVBit 1.7.5, and NVIDIA Compute Sanitizer 2025.2.0.0. All fuzzing harnesses are statically compiled using NVIDIA CUDA compiler (nvcc) 12.9.36 with clang 14 as the host compiler. For closed-source libraries (lower half of \cref{tab:cufuzz-benchmarks}), we statically link fuzzing harnesses with the official pre-built versions shown in the table. 

\subsection{Infrastructure} \label{subsec:methodology-infrastructure}
We run experiments on two servers, each equipped with Intel(R) Xeon(R) Platinum 8362 CPU (64 cores, 2 threads per core) and \SI{1008}{GB} memory. Each server contains eight NVIDIA A40 GPUs with \SI{48}{GB} memory each. The servers run Ubuntu 22.04-x86\_64-standard-uefi with NVIDIA driver 570.144 and CUDA 12.9 toolkit. 

To minimize variability across runs, we set CPU clock speed to \SI{3.2}{GHz} and GPU logic and memory clock speeds to \SI{1740}{MHz} and \SI{7200}{MHz}, respectively. We use separate Docker containers for each benchmark, with each container having exclusive access to 16 CPU cores, \SI{120}{GB} memory, and a single A40 GPU. We run exactly one fuzzing harness per GPU to avoid context switching. 

\subsection{Fuzzing Configurations} \label{subsec:methodology-configurations}
To evaluate our tool's effectiveness, we run and compare the following configurations: 
\begin{itemize}
	\item \textit{AFL++}: baseline approach with only host-side coverage collection enabled (through ``afl-clang-fast'' compiler instrumentation).
	\item \textit{\pname{}}: our comprehensive approach including AFL++ plus device-side coverage collection (through NVBit) and full sanitization (through Google's AddressSanitizer and NVIDIA Compute Sanitizer's memcheck, racecheck, and initcheck tools).
	\item \textit{\pnameNoDCov{}}: \pname{} variant with device-side coverage collection disabled.
	\item \textit{\pnameNoSan{}}: \pname{} variant with sanitization disabled (only host- and device-side coverage collection enabled).
	\item \textit{\pnamePersistent{}}: \pname{} variant with persistent mode enabled, including device-side coverage collection and sanitization.
\end{itemize}

We run these five configurations (plus three more described in~\cref{subsec:cufuzz-sanitization-modes}) on our 14 benchmarks three times for 24 hours each, totaling 8,064 GPU hours, and report the best per-configuration result across the three runs. We use the same initial seed corpus per benchmark for all configurations. Seed counts range from 1 to 5, depending on each benchmark and its input format. 

For sanitizer-enabled configurations (\pname{}, \pnameNoDCov{}, and \pnamePersistent{}), we feed all fuzzer-generated inputs to the fuzzing harness and only a subset of these inputs to the four sanitizers. The inputs we feed to the sanitizers are the ones that trigger at least one new edge in either host or device-side code. We compare different strategies for selecting the inputs to feed to the sanitizers in \cref{subsec:cufuzz-sanitization-modes}. 

\section{Evaluation} \label{sec:eval}

This section describes our~\pname{} evaluation. We first present our research questions, then detail the experimental results. Finally, we analyze our experimental results and answer each research question. Our evaluation was guided by the following research questions:
\begin{itemize}
	\item \textbf{RQ1: Bug finding.} How effective is \pname{} in finding bugs in the target libraries?
	\item \textbf{RQ2: Coverage.} How does \pname{}'s coverage compare to baseline AFL++ fuzzer coverage?
	\item \textbf{RQ3: Performance.} How is \pname{}'s performance impacted by its individual components?
	\item \textbf{RQ4: Persistent mode.} Does persistent mode improve \pname{}'s performance?
	\item \textbf{RQ5: Sanitization.} How does input selection for sanitization impact \pname{}'s coverage?
	\item \textbf{RQ6: Efficiency.} Which \pname{} configuration provides the best time to exposure of bugs?
	\item \rev{\textbf{RQ7: Kernel-level fuzzing.} How does \pname{} compare to kernel-level fuzzing approaches?}
\end{itemize}

\subsection{Bug Finding} \label{subsec:bug-finding}
A fuzzer's most important metric is its ability to find bugs in tested programs. We assess \pname{}'s bug-finding capability by running it for 24 hours on target programs from \cref{tab:cufuzz-benchmarks}. \cref{tab:cufuzz-bugs} summarizes bugs found by \pname{}, including target library, bug type, file or kernel name, bug category (host or device), \rev{used sanitizer}, and status (pending, confirmed, fixed). We record issues as ``confirmed'' only when developers can reproduce them and as ``fixed'' when a pull request containing the bug fix is merged into the main branch as a result of \pname{}'s findings. 

\begin{table}[!ht]
    \centering
    \caption{List of bugs found by \pname{}.}
	\resizebox{0.98\textwidth}{!}{
    \label{tab:cufuzz-bugs}
    \begin{tabular}{lllllll}
        \hline
        Bug\# & Target library & Bug type & File/kernel name & Category & \rev{Sanitizer} & Status \\
        \hline
		1 & attention & Heap buffer overflow & reference.h:13 & host & \rev{ASan} & fixed \\
		2 & boxfilter & Invalid global read of size 4 bytes & col\_kernel & device & \rev{memcheck} & fixed \\
		3 & boxfilter & Uninitialized global read of size 4 bytes & row\_kernel & device & \rev{initcheck} & fixed \\
		4 & boxfilter & Uninitialized global read of size 4 bytes & col\_kernel & device & \rev{initcheck} & fixed \\
		5 & boxfilter & Heap buffer overflow & reference.cpp:121 & host & \rev{ASan} & fixed \\
		6 & crs & Floating-point exception & main.cu:81 & host & \rev{crash} & fixed \\
		7 & dxtc2 & Invalid global read of size 4 bytes & compress & device & \rev{memcheck} & fixed \\
		8 & dxtc2 & Invalid global write of size 8 bytes & compress & device & \rev{memcheck} & fixed \\
		9 & dxtc2 & Data race in shared memory & compress & device & \rev{racecheck} & fixed \\
		10 & dxtc2 & Heap buffer overflow & shrUtils.cu:1122 & host & \rev{ASan} & fixed \\
		11 & lud & Invalid global read of size 4 bytes & lud\_diagonal & device & \rev{memcheck} & fixed \\
		12 & lud & Stack buffer overflow & common/common.cu:161 & host & \rev{ASan} & fixed \\
		13 & medianfilter & Invalid global write of size 4 bytes & ckMedian & device & \rev{memcheck} & fixed \\
		14 & medianfilter & Uninitialized global read of size 4 bytes & ckMedian & device & \rev{initcheck} & fixed \\
		15 & medianfilter & Data race in shared memory & ckMedian & device & \rev{racecheck} & fixed \\
		16 & medianfilter & Heap buffer overflow & MedianFilterHost.cu:70 & host & \rev{ASan} & fixed \\
		17 & medianfilter & Heap buffer overflow & MedianFilterHost.cu:85 & host & \rev{ASan} & fixed \\
		18 & recursiveGaussian & Heap buffer overflow & shrUtils.cu:1302 & host & \rev{ASan} & fixed \\
		19 & recursiveGaussian & Invalid global read of size 4 bytes & RecursiveRGBA & device & \rev{memcheck} & fixed \\
		20 & recursiveGaussian & Uninitialized global read of size 4 bytes & RecursiveRGBA & device & \rev{initcheck} & fixed \\
		21 & recursiveGaussian & Uninitialized global read of size 4 bytes & Transpose & device & \rev{initcheck} & fixed \\
		22 & recursiveGaussian & Heap buffer overflow & RecursiveGaussianHost.cu:180 & host & \rev{ASan} & fixed \\
		23 & seam-carving & Invalid global read of size 4 bytes & compute\_costs\_kernel & device & \rev{memcheck} & fixed \\
		24 & urng & Heap buffer overflow & SDKBitMap.h:368 & host & \rev{ASan} & fixed \\
		\hline
		25 & nvTIFF & Invalid global write of size 1 bytes & batchedCopyLittleEndian\_k & device & \rev{memcheck} & fixed	\\
		26 & nvTIFF & Invalid global write of size 2 bytes & reshapeStrilesRGBInterleaved\_k & device & \rev{memcheck} & fixed	\\
		27 & nvTIFF & Floating-point exception & StripedTiffImageFile & host & \rev{crash} & fixed	\\  
		28 & nvTIFF & Race condition in shared memory & batchedLZWDecompress\_k & device & \rev{racecheck} & fixed \\  
		29 & nvTIFF & Uninitialized global read of size 1 byte & compressStrips\_k & device & \rev{initcheck} & fixed \\  
		30 & nvJPEG & Segmentation fault & libnvjpeg\_static & host & \rev{crash} & fixed	\\  
		31 & nvJPEG & Invalid global write of size 1 bytes & ycbcr\_to\_format\_kernel\_roi & device & \rev{memcheck} & fixed	\\
		32 & nvJPEG2000 & Invalid global write of size 1 bytes & idwt & device & \rev{memcheck} & fixed	\\
		33 & nvJPEG2000 & Invalid global write of size 1 bytes & lossy\_mct\_levelshift & device & \rev{memcheck} & fixed	\\
		34 & nvJPEG2000 & Invalid global write of size 1 bytes & lossless\_mct\_levelshift & device & \rev{memcheck} & fixed	\\
		35 & nvJPEG2000 & Uninitialized global read of size 4 bytes & idwt & device & \rev{initcheck} & fixed	\\
		36 & nvJPEG2000 & Uninitialized global read of size 4 bytes & inv\_quantize\_partial & device & \rev{initcheck} & fixed	\\
		37 & nvJPEG2000 & Uninitialized global read of size 4 bytes & tier1decodemultiple\_channels\_k & device & \rev{initcheck} & fixed	\\
		38 & nvJPEG2000 & Uninitialized global read of size 4 bytes & htTier1MelAndVlcDecodeMultiple & device & \rev{initcheck} & fixed	\\
		39 & nvJPEG2000 & Race condition in shared memory & tier1decodemultiple\_channels\_k & device & \rev{racecheck} & fixed	\\
		40 & nvJPEG2000 & Segmentation fault & tier2Decode::decodeCPRL & host & \rev{crash} & fixed	\\  
		41 & cuDNN & Race condition in shared memory & scudnn\_winograd\_128x128 & device & \rev{racecheck} & confirmed	\\
		42 & cuDNN & Uninitialized global read of size 4 bytes & conv2d\_grouped\_direct\_kernel & device & \rev{initcheck} & confirmed	\\
		43 & cuDNN & Uninitialized global read of size 4 bytes & implicit\_convolve\_sgemm & device & \rev{initcheck} & confirmed	\\
        \hline
    \end{tabular}
	}
    \Description{List of bugs found by \pname{}.}
\end{table}

\begin{figure}[!ht] 
	\centering
	\includegraphics[width=0.9\textwidth]{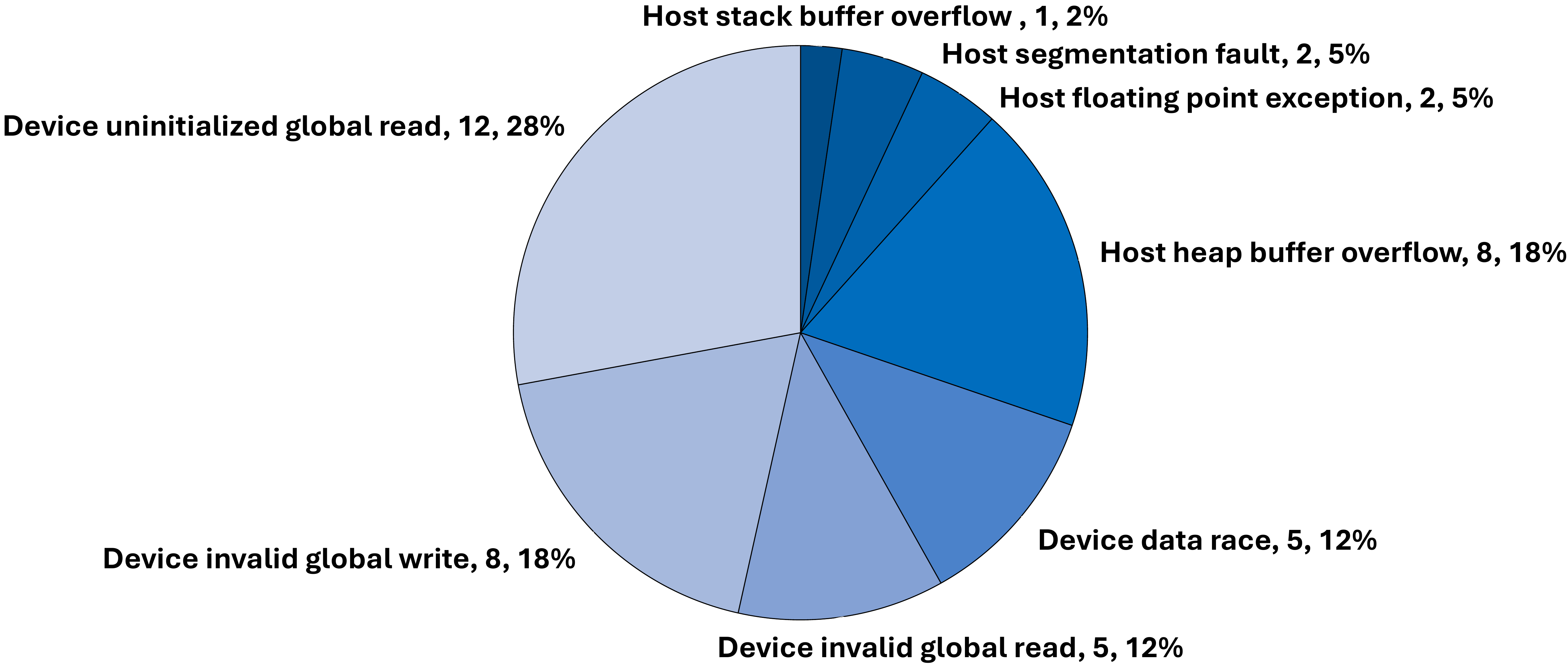}
	\caption{Distribution of bug types found by~\pname{}.} \label{fig:cufuzz-bugs} 
	\Description{Distribution of bug types found by~\pname{}.}
\end{figure}

\cref{fig:cufuzz-bugs} shows the distribution of bug types found by \pname{}. \pname{} discovers a wide range of bugs in both host- and device-side code: 8 heap buffer overflows, 1 stack buffer overflow, 2 floating-point exceptions, 2 segmentation faults, 13 out-of-bounds device accesses, 5 shared memory data races, and 12 uninitialized device reads. Notably, 19 of 43 bugs occur in commercial libraries, highlighting \pname{}'s effectiveness in bug detection. Next, we analyze some of the bugs found by~\pname{} in more detail. 

\begin{figure}[!h]
	\centering
	\begin{minipage}[t]{0.47\textwidth}
	\begin{lstlisting}[language=C, caption={Bug \#2 memcheck report.}, label={lst:bug-2-memcheck}, basicstyle=\tiny]
========= Invalid __global__ read of size 4 bytes
=========     at col_kernel(const unsigned int *, unsigned int *, unsigned int, unsigned int, int, float)+0x490
=========     by thread (1,0,0) in block (0,0,0)
=========     Access at 0x7f71b9e01360 is out of bounds
=========     and is 1 bytes after the nearest allocation at 0x7f71b9e00a00 of size 2400 bytes
=========     Saved host backtrace up to driver entry point at kernel launch time
=========         Host Frame: cudaLaunchKernel [0x86c87] in main
=========         Host Frame: BoxFilterGPU(uchar4*, unsigned int*, unsigned int*, unsigned int, unsigned int, int, float, int) in main.cu:187 [0xbe55] in main
=========         Host Frame: main in main.cu:236 [0xb23c] in main
	\end{lstlisting}
	\end{minipage}\hfill
	\begin{minipage}[t]{0.47\textwidth}
		
	\begin{lstlisting}[language=C++, caption={Bug \#2 location in col\_kernel function.}, label={lst:bug-2-col-kernel}, basicstyle=\tiny, escapeinside={(*@}{@*)}]
__global__ void col_kernel (
    const uint* __restrict__ uiSource, uint* __restrict__ uiDest,
    const int uiWidth,
    const int uiHeight,
    const int iRadius,
    const float fScale)
{
    int globalPosX = blockIdx.x * blockDim.x + threadIdx.x;(*@\label{line:bug2-line8}@*)
   (*@\colorbox{green!20}{ if (globalPosX >= uiWidth) return; // the added fix}\label{line:bug2-line9}@*)

    const uint* uiInputImage = &uiSource[globalPosX]; // bug here(*@\label{line:bug2-line11}@*)
    uint* uiOutputImage = &uiDest[globalPosX];
    // ...
	\end{lstlisting}
	\end{minipage}
	
	\caption{Bug \#2 (invalid global read in col\_kernel from \texttt{boxfilter}) root cause analysis.}
	\label{fig:bug-2-boxfilter}
    \Description{Bug \#2 (invalid global read in col\_kernel from \texttt{boxfilter}) root cause analysis.}
    \vspace{-0.1in}
\end{figure}

\subsubsection{Bug Sample (Id \#2)} 
Bug \#2 is an invalid device-side global-memory read in the col\_kernel function of the \texttt{boxfilter} benchmark. This benchmark accepts a ppm image as input and performs simple arithmetic operations on pixel values. \cref{lst:bug-2-memcheck} shows the error reported by Compute Sanitizer's memcheck tool. As shown in \cref{lst:bug-2-col-kernel}, the root cause is that col\_kernel accesses the input image buffer (\hyperref[line:bug2-line11]{line~11}) using a global index (\hyperref[line:bug2-line8]{line~8}) without checking if the index is within buffer bounds. The fix adds a bounds check (\hyperref[line:bug2-line9]{line~9}) ensuring only threads with indices smaller than the image width access the image buffer.

\begin{figure}[!h]
	\centering
	\begin{minipage}[t]{0.49\textwidth}
	\begin{lstlisting}[language=C, caption={Bug \#23 memcheck report.}, label={lst:bug-23-memcheck}, basicstyle=\tiny]
========= Program hit cudaErrorMemoryAllocation (error 2) due to "out of memory" on CUDA API call to cudaMalloc.
=========     Saved host backtrace up to driver entry point at error
=========         Host Frame: main [0x9208] in main
========= 
========= Program hit cudaErrorInvalidValue (error 1) due to "invalid argument" on CUDA API call to cudaMemcpy.
=========     Saved host backtrace up to driver entry point at error
=========         Host Frame: main [0x8d9d] in main
========= 
========= Invalid __global__ read of size 4 bytes
=========     at compute_costs_kernel(const uchar4 *, short *, short *, short *, int, int, int)+0x110
=========     by thread (1,4,0) in block (0,0,0)
=========     Access at 0x3fc84 is out of bounds
=========     and is 140660749108092 bytes before the nearest allocation at 0x7fee22000000 of size 1137491208 bytes
=========     Saved host backtrace up to driver entry point at kernel launch time
=========         Host Frame: main [0x92fa] in main
=============================================
	\end{lstlisting}
	\end{minipage}\hfill
	\begin{minipage}[t]{0.49\textwidth}
		
	\begin{lstlisting}[language=C++, caption={Bug \#23 location.}, label={lst:bug-23-compute-costs-kernel}, basicstyle=\tiny, escapeinside={(*@}{@*)}]
#define CUDA_CHECK(call)                                                    \(*@\label{line:bug23-line1}@*)
do {                                                                        \
    cudaError_t err_ = call;                                                \
    if (err_ != cudaSuccess) {                                              \
        fprintf(stderr, "CUDA error at %s:%d code=%d(%s) \"%s\" \n",        \
                __FILE__, __LINE__, err_, cudaGetErrorString(err_), #call); \
        exit(EXIT_FAILURE);                                                 \
    }                                                                       \
} while (0)(*@\label{line:bug23-line9}@*)
// ...
(*@\colorbox{red!20}{cudaMalloc((void**)\&d\_pixels, img\_bytes);}\label{line:bug23-line11}@*)
(*@\colorbox{red!20}{cudaMemcpy(d\_pixels, h\_pixels, img\_bytes}\label{line:bug23-line12}@*)
(*@\colorbox{red!20}{     , cudaMemcpyHostToDevice);}@*)
(*@\colorbox{green!20}{CUDA\_CHECK(cudaMalloc((void**)\&d\_pixels, img\_bytes));}@*)
(*@\colorbox{green!20}{CUDA\_CHECK(cudaMemcpy(d\_pixels, h\_pixels, img\_bytes }@*)
(*@\colorbox{green!20}{    , cudaMemcpyHostToDevice));}@*)
// ...
	\end{lstlisting}
	\end{minipage}
	\caption{Bug \#23 (invalid global read in compute\_costs\_kernel from \texttt{seam-carving}) root cause analysis.}
	\label{fig:bug-23-seam-carving}
    \Description{Bug \#23 (invalid global read in compute\_costs\_kernel from \texttt{seam-carving}) root cause analysis.}
\end{figure}

\subsubsection{Bug Sample (Id \#23)}
Bug \#23 is another case of invalid device-side global memory reads. It impacts the compute\_costs\_kernel function of the \texttt{seam-carving} benchmark. This benchmark accepts a jpg image as input and performs the seam carving algorithm. \cref{lst:bug-23-memcheck}, which shows the error reported by Compute Sanitizer's memcheck tool, reveals that the invalid access is in fact preceded by a memory allocation error (out-of-memory) in cudaMalloc. The problem is that the original code (\hyperref[line:bug23-line11]{lines~11-13} in \cref{lst:bug-23-compute-costs-kernel}) never checks CUDA API return values at runtime. As a result, the program continues execution and launches device-side kernels even after an API failure. \rev{This bug manifests only with large input images that request allocations exceeding available GPU memory, causing \texttt{cudaMalloc} to fail silently.} One potential fix is to encapsulate all CUDA API calls with checks (\hyperref[line:bug23-line1]{lines~1--9}) to ensure proper application exit if any CUDA API call fails. This bug demonstrates the importance of considering host- and device-side code interaction when identifying and fixing bugs.
	
\begin{figure}[!h]
	\centering
	\begin{minipage}[t]{0.49\textwidth}
	\begin{lstlisting}[language=C, caption={Bug \#24 address sanitizer report.}, label={lst:bug-24-address-sanitizer}, basicstyle=\tiny]
==8121==ERROR: AddressSanitizer: heap-buffer-overflow on address 0x7fcfffdff800 at pc 0x55c89ad345cd bp 0x7ffe7e99f860 sp 0x7ffe7e99f858
READ of size 1 at 0x7fcfffdff800 thread T0
    #0 0x55c89ad345cc in SDKBitMap::load(char const*) src/urng-cuda/../include/SDKBitMap.h:368:39
    #1 0x55c89ad30e44 in main src/urng-cuda/main.cu:38:13

0x7fcfffdff800 is located 0 bytes to the right of 786432-byte region [0x7fcfffd3f800,0x7fcfffdff800)
allocated by thread T0 here:
    #0 0x55c89ad2e69d in operator new[](unsigned long) (src/urng-cuda/main+0xe369d) (BuildId: 19d629d31da22263bff3ab5f79c6f821c59d7085)
    #1 0x55c89ad3353d in SDKBitMap::load(char const*) src/urng-cuda/../include/SDKBitMap.h:320:28

SUMMARY: AddressSanitizer: heap-buffer-overflow src/urng-cuda/../include/SDKBitMap.h:368:39 in SDKBitMap::load(char const*)
	\end{lstlisting}
	\end{minipage}\hfill
	\begin{minipage}[t]{0.49\textwidth}
		
	\begin{lstlisting}[language=C++, caption={Bug \#24 location.}, label={lst:bug-24-sdkbitmap-h}, basicstyle=\tiny, escapeinside={(*@}{@*)}]
void load(const char * filename){
    //...
    // Allocate buffer to hold all pixels
    unsigned int sizeBuffer = size - offset;
    (*@\colorbox{green!20}{if (width * height * (bitsPerPixel \/ 8) != sizeBuffer) \{}\label{line:bug24-line5}@*)
    (*@\colorbox{green!20}{		printf("This is not a valid bitmap file.\\n");}@*)
    (*@\colorbox{green!20}{		fclose(fd); return;}@*)
    (*@\colorbox{green!20}{\}}\label{line:bug24-line8}@*)
    unsigned char * tmpPixels = new unsigned char[sizeBuffer];
    if (tmpPixels == NULL)
    {
        delete colors_;
        colors_ = NULL;
        fclose(fd);	return;
    }
    // Read pixels from file, including any padding
    val = fread(tmpPixels, sizeBuffer * sizeof(unsigned char), 1, fd);
	\end{lstlisting}
	\end{minipage}
	
	\caption{Bug \#24 (heap-buffer-overflow in  SDKBitMap::load from \texttt{urng}) root cause analysis.}
	\label{fig:bug-24-urng}
    \Description{Bug \#24 (heap-buffer-overflow in  SDKBitMap::load from \texttt{urng}) root cause analysis.}
    \vspace{-0.1in}
\end{figure}

\subsubsection{Bug Sample (Id \#24)}
Bug \#24 is a heap-buffer overflow in the SDKBitMap::load function of the \texttt{urng} benchmark. This benchmark loads an input ppm image and generates uniform random noise. \cref{lst:bug-24-address-sanitizer} shows the error reported by Google's AddressSanitizer for the out-of-bounds read. The root cause is that the original code (\cref{lst:bug-24-sdkbitmap-h}) assumes the ``size'' and ``offset'' fields read from the input image header are trusted. Thus, it never compares these fields with the actual input image size. The fix is to add a validation check (\hyperref[line:bug24-line5]{lines~5--8}) ensuring the application exits properly if the input image is not a valid bitmap file. This bug demonstrates \pname{}'s versatility in catching bugs using different sanitizers.

\begin{takeawaybox}
\textit{\textbf{Answer to RQ1:} \pname{} discovered 43 previously-unknown bugs across the tested programs, including 19 in commercial libraries, demonstrating its bug-finding capability}.
\end{takeawaybox}

\subsection{Coverage} \label{subsec:coverage}
\begin{figure}[!h] 
	\centering
	\includegraphics[width=0.98\textwidth]{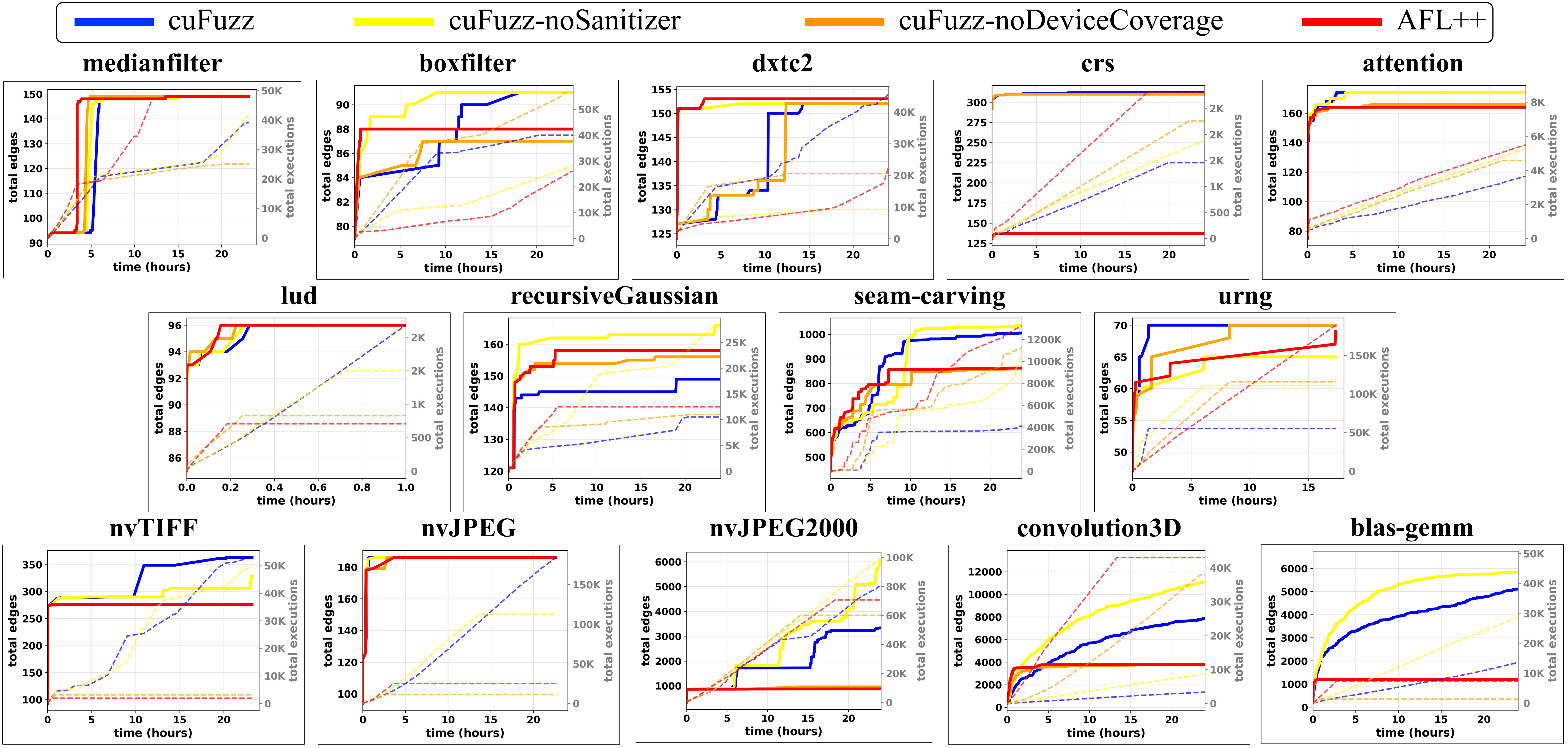}
    \caption{Host- and device-side edge coverage over 24 hours for all 14 benchmarks. \rev{Solid lines show cumulative edges discovered; dashed lines show total executions.}} \label{fig:cufuzz-edges}     
	\Description{\pname{} edge coverage.}
\end{figure}

\begin{figure}[!h] 
	\centering
	\includegraphics[width=0.98\textwidth]{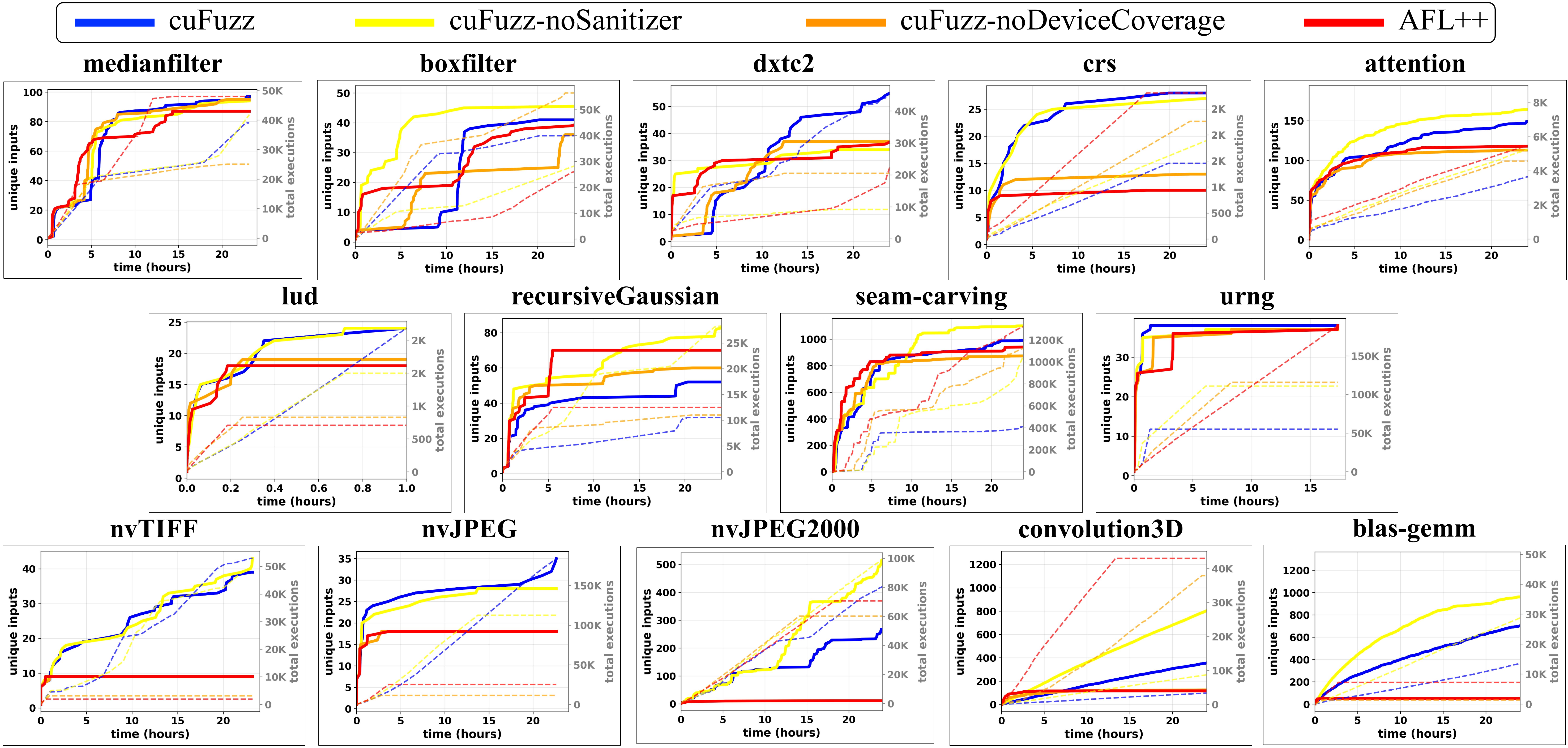}
	\caption{Unique inputs discovered over 24 hours for all 14 benchmarks. \rev{An input is unique if it triggers new edges or crosses hit-count bucket boundaries.}} \label{fig:cufuzz-inputs} 
	\Description{\pname{} unique input coverage.}
\end{figure}

Coverage is the second key fuzzer evaluation metric. It poses the following question: starting with few input seeds covering limited library portions, how deeply can the fuzzer explore? More precisely, how many new edges can a fuzzer discover in a given time period? We evaluate \pname{}'s coverage by comparing its edge and unique input coverage to baseline AFL++. \cref{fig:cufuzz-edges} and \cref{fig:cufuzz-inputs} compare edge and unique input coverage of different \pname{} configurations from \cref{subsec:methodology-configurations} to vanilla AFL++ across all benchmarks. An input is considered unique if it reaches new edges or increases hit counts \rev{enough to cross logarithmic bucket boundaries: AFL++'s host-side buckets (1, 2, 3, 4--7, 8--15, 16--31, 32--127, 128+) or our device-side buckets (1, 2, 3+, 512+, 4096+, 16384+, 65536+) which use coarser granularity to accommodate GPU thread counts.}

To ensure fair comparison, we first run all configurations for the same duration and collect interesting findings in corresponding queues. We then rerun all queue entries per configuration using a modified ``afl-showmap'' tool to collect edge coverage information from both host and device sides (even for configurations not using device-side information like AFL++). The hypothesis is that even AFL++ might accidentally hit device-side edges that need quantification, though they are not part of fuzzer feedback. In \cref{fig:cufuzz-edges} and \cref{fig:cufuzz-inputs}, we show edge and unique input coverage on the main y-axis (solid lines) and total executions on the secondary y-axis (dashed lines).  

Our results reveal several key observations. First, for benchmarks with available source code, vanilla AFL++ achieves better coverage (or reaches the same coverage faster) than \pname{}. This stems from two reasons: (a) AFL++ runs faster without sanitization and device-side coverage slowdowns, and (b) host-side coverage suffices for these benchmarks since they rely on simple kernels where host-side edges indicate device-side execution well. Second, for benchmarks invoking closed-source libraries, \pname{} achieves better coverage than AFL++ due to NVBit benefits. Third, coverage of \pname{} and \pnameNoSan{} is nearly identical with only small delays from sanitizer slowdowns. The same observation applies to AFL++ and \pnameNoDCov{}. Finally, while a single \pname{} execution is slower than AFL++'s single execution, \pname{} achieves the same coverage as AFL++ with fewer executions. This occurs because \pname{} leverages device-side coverage to guide the fuzzing process, generating more effective mutations in fewer executions. 

\begin{figure}[!h] 
	\centering
	\includegraphics[width=0.98\textwidth]{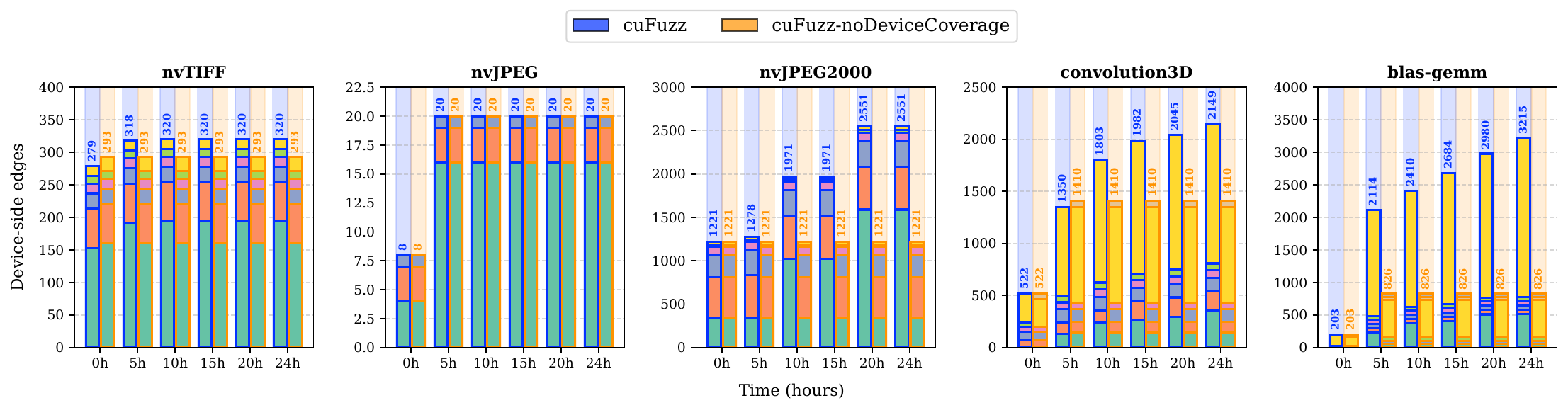}
	\caption{\rev{Device-side edge coverage over time for closed-source libraries, comparing \pname{} (left bars, blue labels) vs.\ \pnameNoDCov{} (right bars, orange labels). Numbers above bars indicate total device-side edges at each time point.}} \label{fig:cufuzz-device-edges} 
	\Description{Device-side edge coverage decomposed by kernel for closed-source libraries.}
\end{figure}

\rev{To further confirm our observation regarding closed-source libraries, \cref{fig:cufuzz-device-edges} reports device-side edge progress during fuzzing for the five closed-source applications, comparing \pname{} with \pnameNoDCov{}. Each stacked bar shows the cumulative device-side edges discovered, decomposed by kernel. As shown, leveraging device-side coverage through NVBit enables \pname{} to discover significantly more device-side edges compared to \pnameNoDCov{}, with improvements ranging from 9\% (\texttt{nvTIFF}) to 289\% (\texttt{blas-gemm}). Open-source benchmarks are omitted since both configurations reach the same edges (per~\cref{fig:cufuzz-edges}). Closed-source libraries benefit more because their pre-compiled binaries lack host-side instrumentation.}

\begin{takeawaybox}
\textit{\textbf{Answer to RQ2:} Enabling device-side coverage in \pname{} ({\color{blue}blue} and {\color{yellow}yellow}) yields substantially higher coverage than baseline AFL++ ({\color{red}red}) for closed-source libraries, \rev{because their pre-compiled binaries lack host-side instrumentation and device-side coverage via NVBit is essential for guiding exploration.} Open-source programs show similar coverage across configurations since host-side instrumentation suffices to guide device-side exploration}.
\end{takeawaybox}

\subsection{Performance} \label{subsec:perf-eval}
\begin{figure}[!h]
	\centering
	\subfloat[Using all inputs.]{\includegraphics[width=0.98\textwidth]{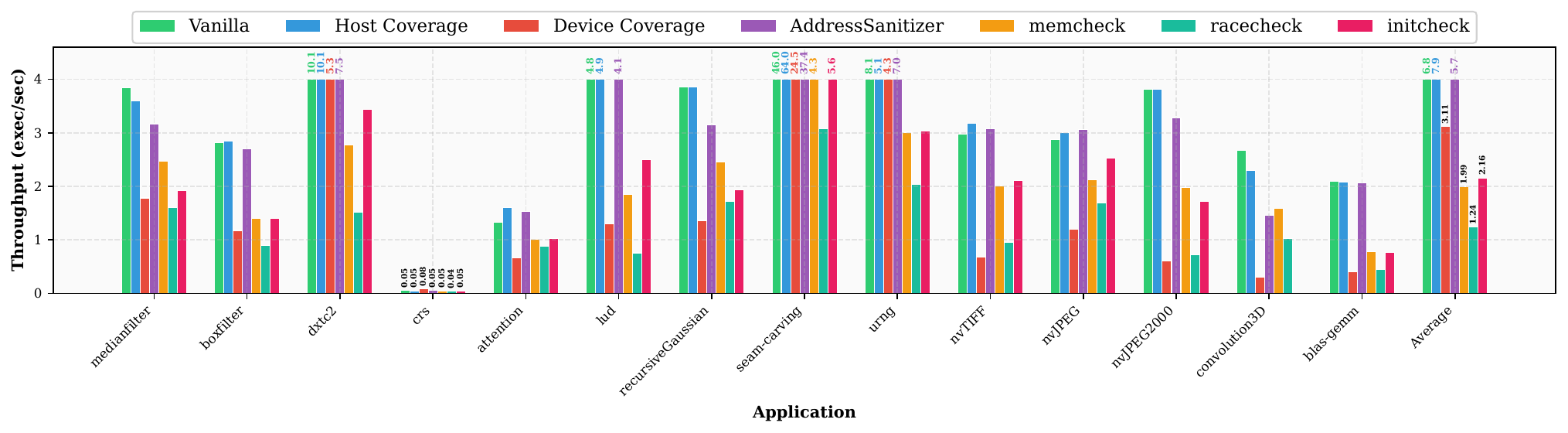}\label{fig:cufuzz-throughput-all-inputs}}\\
	
	\subfloat[Using device-triggering inputs only.]{\includegraphics[width=0.98\textwidth]{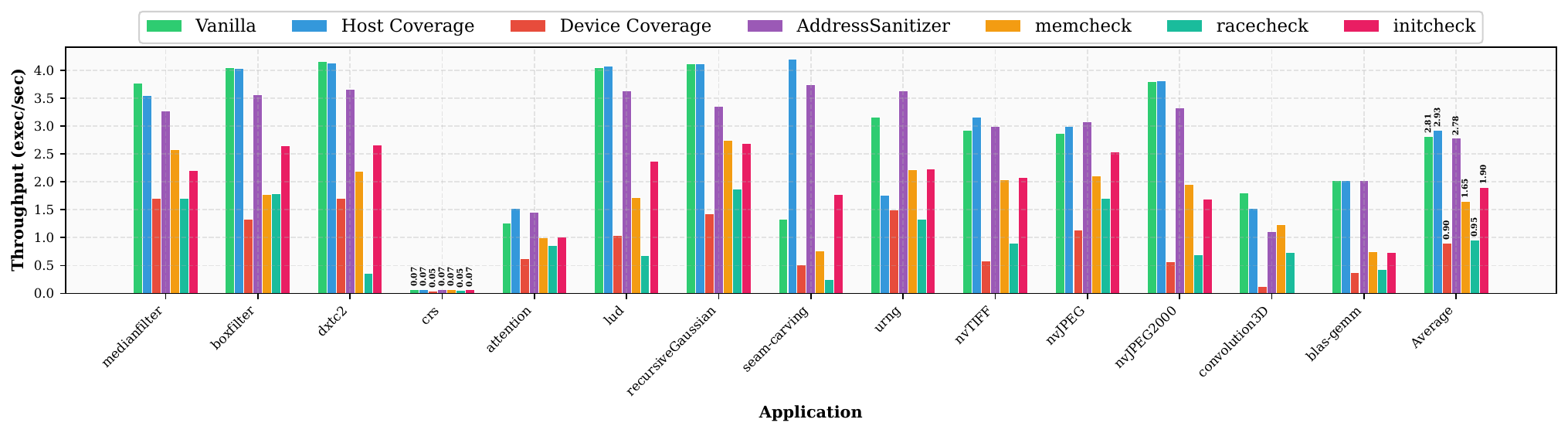}\label{fig:cufuzz-throughput-device-inputs}}
	
	\caption{Throughput \rev{(executions/second)} of individual \pname{} components measured using 100 unique inputs per benchmark. \rev{(a)~All inputs; (b)~Device-triggering inputs only.}}
	\label{fig:cufuzz-throughput-comparison}
	\Description{Throughput of \pname{} components using all inputs and device-triggering inputs.}
\end{figure}

To understand the performance impact of different \pname{} components, we perform an ablation study. We select 100 unique generated inputs per benchmark and run them using ``hyperfine''~\cite{Peter_hyperfine_2023} on individual \pname{} components, reporting median throughput per second (1000/execution time). The goal is to identify speed-of-light throughput when each component is disjointly enabled over the same input set, since comparing throughput across different fuzzing runs introduces significant randomness. \cref{fig:cufuzz-throughput-comparison} shows our ablation study results. 

Considering all 100 inputs regardless of execution pattern (\cref{fig:cufuzz-throughput-all-inputs}), we observe comparable average throughput for vanilla execution, host-side coverage, and AddressSanitizer. This is expected since these three variants rely on compile-time instrumentation with similarly low overhead. Conversely, we observe 55\% throughput reduction for device-side coverage due to NVBit's runtime overheads. Similarly, we observe 71\% throughput reduction for memcheck and 82\% for racecheck due to Compute Sanitizer's runtime overheads. 

While average results are expected, \cref{fig:cufuzz-throughput-all-inputs} contains outliers where vanilla throughput significantly exceeds other workloads. For example, \texttt{seam-carving}'s vanilla throughput is \SI{46}{exec/sec} (\SI{24}{exec/sec} with NVBit). To understand this behavior, we investigated per-input throughput and noticed a striking difference between inputs rejected early by host-side code and inputs triggering at least one kernel execution. 

For device-triggering inputs, \texttt{seam-carving} vanilla throughput is only \SI{1.2}{exec/sec} (\SI{0.5}{exec/sec} with NVBit), comparable to other workloads. We define device-triggering inputs as the inputs that actually manage to execute at least one device-side kernel. Other inputs were too shallow, failing host-side sanity checks and causing early program termination. This explains why vanilla and NVBit throughput were notably high when no device-side execution occurred. Conversely, Compute Sanitizer tools always incur the same overhead regardless of input, likely due to runtime loading and initialization costs of the sanitizer itself. We report only device-triggering input throughput in \cref{fig:cufuzz-throughput-device-inputs}. On average, NVBit reduces throughput by 67\% and memcheck by 39\%, while host-side sanitization has negligible throughput impact. 
	
\begin{takeawaybox}
\textit{\textbf{Answer to RQ3:} Device-side coverage collection is \pname{}'s primary throughput bottleneck, reducing throughput by 67\% for inputs that trigger device execution. The throughput impact of device-side sanitization varies with the specific tool enabled}.
\end{takeawaybox}

\subsection{Persistent Mode} \label{subsec:persistent-eval}
\begin{figure}[!h] 
	\centering
	\includegraphics[width=0.98\textwidth]{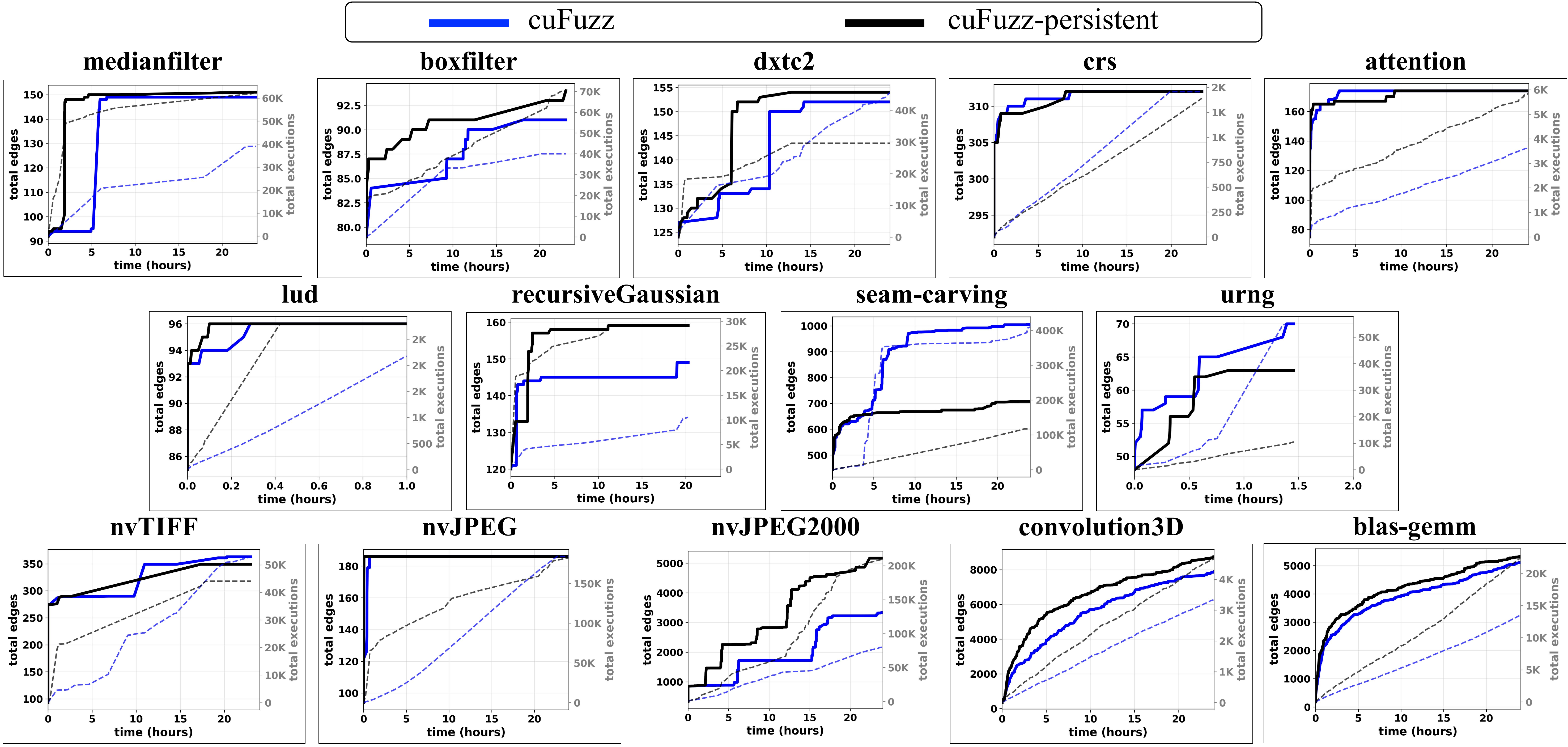}
    \caption{\rev{Edge coverage comparison: \pnamePersistent{} (black) vs.\ regular \pname{} (blue) over 24 hours.}} \label{fig:cufuzz-edges-persistent}  
	\Description{\pname{} edge coverage in persistent mode.}
\end{figure}

To understand persistent mode's impact on \pname{}, \cref{fig:cufuzz-edges-persistent} compares \pname{}'s edge coverage in persistent mode (black line) versus regular single-input-per-process \pname{} from \cref{fig:cufuzz-edges} (blue line). We observe that \pnamePersistent{} achieves higher coverage than \pname{} in the same time for seven applications, while both approaches achieve the same coverage for four applications. For the remaining three programs (\texttt{seam-carving}, \texttt{urng}, and \texttt{nvTIFF}), \pnamePersistent{} does not improve fuzzing throughput \rev{because the harness requires expensive reinitialization inside the loop. For \texttt{nvTIFF} and \texttt{seam-carving}, we must call \texttt{cudaMalloc} and \texttt{cudaFree} for large buffers in every iteration to handle variable-sized inputs and properly reset state between runs. Moving these allocations outside the loop causes crashes when subsequent inputs require different buffer sizes. For \texttt{urng}, the kernel execution itself is so short-lived (sub-millisecond) that the overhead of the persistent loop bookkeeping negates any benefit from avoiding process restart.}

\begin{takeawaybox}
\textit{\textbf{Answer to RQ4:} In 11 of 14 programs, \pnamePersistent{} delivered equal or better coverage than single-input-per-process~\pname{}, highlighting its potential}.
\end{takeawaybox}

\subsection{Sanitization} \label{subsec:cufuzz-sanitization-modes}
\begin{figure}[!ht] 
	\centering
	\includegraphics[width=0.98\textwidth]{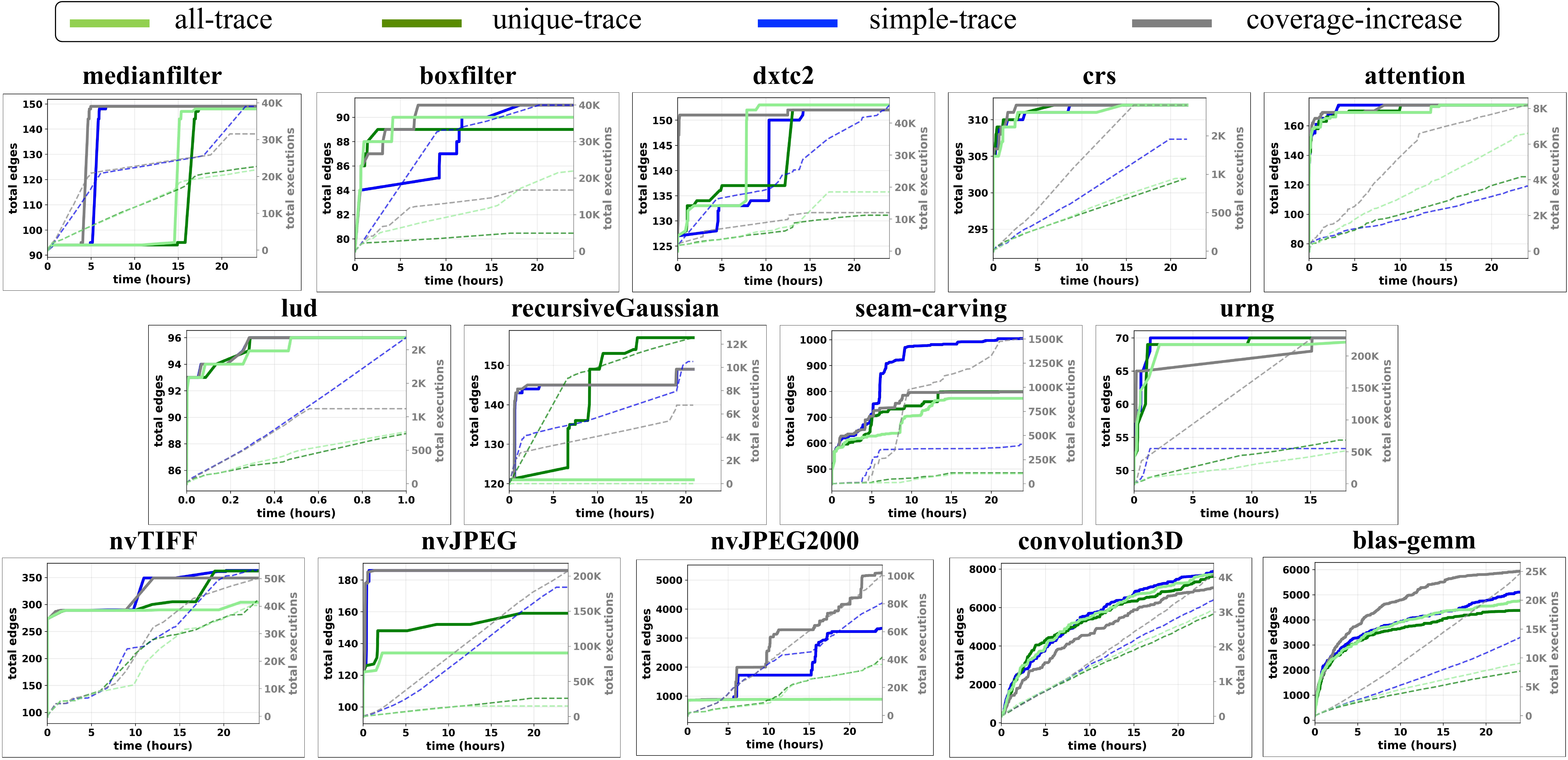}
	\caption{\pname{} edge coverage in various sanitization strategies. ``simple-trace'' is the default \pname{} strategy.} \label{fig:cufuzz-edges-sanitization} 
	\Description{\pname{} edge coverage in various sanitization strategies. ``simple-trace'' is the default \pname{} strategy.}
    \vspace{-0.1in}
\end{figure}

Given Compute Sanitizer tools' large impact on fuzzing throughput (\cref{fig:cufuzz-throughput-comparison}), running all sanitizers on every input would be suboptimal. Inspired by prior work~\cite{Kong2025:SAND}, we evaluate four strategies for subsetting inputs fed to sanitizers: (1) ``all-trace'': runs all sanitizers on every input, (2) ``unique-trace'': runs all sanitizers on inputs with unique execution paths, (3) ``simple-trace'': runs sanitizers on inputs with unique execution paths without considering thread/edge count, and (4) ``coverage-increase'': runs sanitizers on inputs causing fuzzing engine coverage increases. \cref{fig:cufuzz-edges-sanitization} summarizes \pname{}'s edge coverage with each strategy. 

As the most sensitive strategy, all-trace has the lowest performance. While it achieves the same coverage as other strategies for some benchmarks (e.g., \texttt{medianfilter}), it is generally slower. Conversely, coverage-increase is the most performant strategy as the least sensitive one, running only inputs that are part of the fuzzing queue through sanitizers. The unique-trace and simple-trace strategies balance sensitivity and runtime overheads; thus, the latter (simple-trace) is enabled by default in our main experiments (\cref{subsec:methodology-configurations}). We justify this decision in the following section. 

\begin{takeawaybox}
\textit{\textbf{Answer to RQ5:} While all strategies progress similarly over time, the less sensitive input selection methods ``simple-trace'' and ``coverage-increase'' run faster than the others because they incur lower device-sanitization overhead}.
\end{takeawaybox}

\subsection{Time to Exposure} \label{subsec:time-to-exposure}
To understand device-side sanitization's impact on \pname{}'s bug-finding capability and to quantify the effectiveness of the different sanitization strategies, we collected bug exposure times for different \pname{} configurations. \cref{tab:cufuzz-time-to-exposure} shows the results with the shortest exposure time highlighted in blue. After three 24-hour fuzzing experiments, we collect unique bugs found by each configuration. We applied Google's AddressSanitizer and Compute Sanitizer's memcheck, racecheck, and initcheck tools to all crash folder contents, collecting crashes from generated test inputs per configuration. We identified unique crash bugs based on collected error messages, first removing crashes with identical stack traces, then manually identifying crashes with different stack traces but suspected shared root causes. We identify the following observations. 

First, without device-side sanitization (AFL++ and \pnameNoSan{} columns in \cref{tab:cufuzz-time-to-exposure}), the fewest bugs are found (11 and 9 out of 43, respectively). Only host-side (H) bugs causing application-level crashes are captured. Enabling device-side sanitization (even without device-side coverage, such as \pnameNoDCov{}) significantly increases bugs found (30 out of 43). 

Second, comparing sanitization strategies, simple-trace strategy (\pname{} column) finds the most bugs (36 out of 43), followed by coverage-increase strategy (35 out of 43). Despite running every input through sanitizers, all-trace strategy finds only 28 bugs, mainly because the 24-hour limit expired before exploring sufficient paths. Thus, we use simple-trace strategy by default in our main experiments (\cref{subsec:coverage} and~\cref{subsec:persistent-eval}). Finally, enabling persistent mode identifies 35 bugs, including three bugs captured only in \pnamePersistent{} runs. It also has the largest number of bugs with the least time to exposure (16 out of 43). 

\begin{table}[!ht]
    \newcolumntype{Y}{>{\centering\arraybackslash}p{1.5cm}}
    \centering
    \caption{Time to exposure of bugs for different \pname{} configurations.}
    \resizebox{0.98\textwidth}{!}{
    \label{tab:cufuzz-time-to-exposure}
    \begin{tabular}{c c Y Y Y Y Y Y Y Y}
        \hline
        \multicolumn{1}{c}{Bug} & \multicolumn{1}{c}{\multirow{2}{*}{Library}} & \multicolumn{1}{c}{\multirow{2}{*}{\pnameAFL{}}} & \multicolumn{1}{c}{\pname{}-} & \multicolumn{1}{c}{\pname{}-} & \multicolumn{1}{c}{\multirow{2}{*}{\pname{}}} & \multicolumn{1}{c}{\pname{}-} & \multicolumn{1}{c}{\pname{}-} & \multicolumn{1}{c}{\pname{}-} & \multicolumn{1}{c}{\pname{}-} \\
        \multicolumn{1}{c}{id}  &                             &                                 & \multicolumn{1}{c}{noSanitizer} & \multicolumn{1}{c}{noDevCov} & & \multicolumn{1}{c}{persistent} & \multicolumn{1}{c}{all-trace} & \multicolumn{1}{c}{unique-trace} & \multicolumn{1}{c}{cov-increase} \\
        \hline
        1 & attention (H) & 20:57:35 & - & 21:19:16 & - & - & - & - & \textcolor{blue}{04:39:38} \\
        2 & boxfilter (D) & - & - & 00:26:03 & 00:34:30 & \textcolor{blue}{00:00:22} & 00:29:20 & 00:30:31 & 00:36:38 \\
        3 & boxfilter (D) & - & - & 03:27:40 & 04:35:14 & \textcolor{blue}{00:15:15} & 00:38:39 & 00:39:15 & 00:38:21 \\
        4 & boxfilter (D) & - & - & 05:27:18 & 06:00:03 & 01:09:55 & \textcolor{blue}{00:42:34} & 00:44:07 & 06:33:34 \\
        5 & boxfilter (H) & 00:19:36 & 00:33:02 & 04:24:21 & 06:00:03 & \textcolor{blue}{00:15:45} & 00:37:47 & 00:38:26 & 00:37:37 \\
        6 & crs (H) & 00:07:04 & \textcolor{blue}{00:05:35} & 00:07:47 & 00:17:19 & 00:41:32 & 00:15:20 & 00:17:01 & 00:37:54  \\
        7 & dxtc2 (D) & - & - & 08:12:47 & 10:45:04 & 00:49:27 & 07:50:27 & 04:44:39 & \textcolor{blue}{00:04:23}    \\
        8 & dxtc2 (D) & - & - & 10:17:03 & 09:39:31 & 00:51:51 & 07:48:09 & 05:01:39 & \textcolor{blue}{00:04:02} \\
        9 & dxtc2 (D) & - & - & \textcolor{blue}{00:00:33} & \textcolor{blue}{00:00:33} & 00:16:24 & 00:00:34 & \textcolor{blue}{00:00:33} & 00:04:02 \\
        10 & dxtc2 (H) & - & - & 18:52:06 & 12:58:30 & \textcolor{blue}{02:12:57} & 07:48:09 & 20:34:14 & - \\
        11 & lud (D) & - & - & \textcolor{blue}{00:00:04} & 00:00:11 & 00:00:05 & 00:00:17 & 00:00:12 & 00:00:09 \\
        12 & lud (H) & 00:00:38 & 00:00:41 & \textcolor{blue}{00:00:12} & 00:00:48 & 00:00:17 & 00:01:09 & 00:01:01 & 00:00:51 \\
        13 & medianfilter (D) & - & - & \textcolor{blue}{04:54:32} & 06:43:55 & - & 22:31:06 & 15:31:42 & 06:03:30 \\
        14 & medianfilter (D) & - & - & 04:22:53 & 05:26:24 & \textcolor{blue}{01:10:59} & 15:07:54 & 15:13:37 & 04:31:17 \\
        15 & medianfilter (D) & - & - & 04:54:32 & 05:42:48 & - & 15:29:11 & 15:19:32 & \textcolor{blue}{04:36:40} \\
        16 & medianfilter (H) & \textcolor{blue}{02:01:07} & 04:26:50 & 04:54:32 & 05:42:48 & 03:59:49 & 17:28:04 & 15:31:42 & 04:36:40 \\
        17 & medianfilter (H) & 02:31:45 & 04:40:24 & 04:37:16 & 05:53:37 & \textcolor{blue}{01:11:39} & 15:14:51 & 15:19:32 & 04:44:09 \\
        18 & recursiveGaussian (H) & - & - & 06:49:16 & \textcolor{blue}{04:03:00} & - & - & 06:48:36 & 14:43:22	\\
        19 & recursiveGaussian (D) & - & - & - & 02:29:48 & - & - & 06:27:51 & \textcolor{blue}{02:00:43} \\
        20 & recursiveGaussian (D) & - & - & 00:29:34 & 00:31:04 & 00:27:49 & 00:30:28 & \textcolor{blue}{00:26:45} & 00:30:53 \\
        21 & recursiveGaussian (D) & - & - & 01:04:51 & 01:37:08 & - & 13:43:10 & 07:30:35 & \textcolor{blue}{01:04:13}       \\
        22 & recursiveGaussian (H) & 01:11:02 & 01:07:51 & 01:04:51 & 01:37:08 & \textcolor{blue}{00:28:13} & 05:26:07 & 01:04:13 & 01:04:26	\\
        23 & seam-carving (D) & - & - & \textcolor{blue}{04:11:35} & 13:52:39 & - & - & - & -        \\
        24 & urng (H) & \textcolor{blue}{00:02:17} & 00:03:02 & 00:03:21 & 00:03:23 & 00:19:16 & 00:22:36 & 00:32:40 & 00:02:48 \\
        \hline
        25 & nvTIFF (D) & - & - & \textcolor{blue}{00:34:13} & 06:49:08 & 01:01:19 & - & - & - \\
        26 & nvTIFF (D) & - & - & \textcolor{blue}{01:15:46} & - & - & 10:05:52 & 11:02:27 & - \\
        27 & nvTIFF (H) & \textcolor{blue}{00:15:34} & 06:57:32 & 00:33:26 & 06:48:51 & 01:01:00 & - & - & 06:39:07      \\
        28 & nvTIFF (D) & - & - & - & 00:59:35 & 01:15:58 & 01:15:13 & 01:17:39 & \textcolor{blue}{00:53:19} \\
        29 & nvTIFF (D) & - & - & - & - & \textcolor{blue}{11:04:36} & - & - & -  \\
        30 & nvJPEG (H) & \textcolor{blue}{00:08:53} & 00:16:29 & 00:10:37 & 02:10:21 & 00:17:52 & 01:02:40 & 01:55:11 & 01:33:27 \\
        31 & nvJPEG (D) & -  & -  & -  & 00:44:31 & \textcolor{blue}{00:00:02} & 00:04:46 & 00:05:36 & 00:38:50  \\
        32 & nvJPEG2000 (D) & -  & -  & -  & 06:04:46 & \textcolor{blue}{02:06:56} & 13:06:34 & - & 05:50:11 \\
        33 & nvJPEG2000 (D) & -  & -  & -  & 05:41:34 & 12:27:11 & - & - & \textcolor{blue}{05:36:59}       \\
        34 & nvJPEG2000 (D) & -  & -  & -  & - & \textcolor{blue}{12:15:39} & -  & -  & - \\
        35 & nvJPEG2000 (D) & -  & -  & -  & - & \textcolor{blue}{07:25:03} & -  & -  & 09:02:03 \\
        36 & nvJPEG2000 (D) & -  & -  & -  & - & \textcolor{blue}{07:07:57} & -  & -  & -  \\
        37 & nvJPEG2000 (D) & - & - & - & 05:49:55 & \textcolor{blue}{02:10:20} & - & - & 11:54:45 \\
        38 & nvJPEG2000 (D) & -  & -  & -  & 13:15:16 & 12:29:05 & - & - & \textcolor{blue}{06:03:28} \\
		39 & nvJPEG2000 (D) & - & - & - & - & 12:27:11 & - & - & \textcolor{blue}{05:57:24}  \\
        40 & nvJPEG2000 (H) & \textcolor{blue}{00:31:08} & - & 00:38:11 & 06:29:29 & 03:36:02 & - & - & -  \\
        41 & cuDNN (D) & - & - & - & 08:56:58 & \textcolor{blue}{00:25:47} & 03:02:12 & 02:36:43 & 03:29:11 \\
        42 & cuDNN (D)& - & - & 00:31:27 & 00:18:48 & \textcolor{blue}{00:17:52} & 00:39:01 & 00:24:18 & 00:22:12 \\
        43 & cuDNN (D)& - & - & 00:04:31 & 00:02:06 & 00:01:13 & 00:01:52 & \textcolor{blue}{00:01:10} & 00:07:40	\\
        \hline
        & Total findings & 11 & 9 & 30 & 36 & 35 & 28 & 29 & 35 \\
        \hline
    \end{tabular}
    }
    \Description{Time to exposure of bugs for different \pname{} configurations.}
\end{table}

\begin{takeawaybox}
\textit{\textbf{Answer to RQ6:} The ``simple-trace'' \pname{} configuration found the most bugs (83\%), while its persistent-enabled variant achieved the shortest time to exposure for the largest bugs share (37\%)}.
\end{takeawaybox}

\rev{
\subsection{Comparison with Kernel-Level Fuzzing} \label{subsec:kernel-fuzzing}
Prior work on GPU fuzzing~\cite{Li2022:CVFuzz, Peng2020:OpenCLFuzz} proposed kernel-level fuzzing, which isolates individual GPU kernels and permutes their input parameters independently. To understand the trade-offs between kernel-level and whole-program fuzzing, we implemented kernel-level harnesses for 22 kernels across nine open-source benchmarks (closed-source libraries are excluded since kernel names and arguments are hidden). Each harness allocates buffers, initializes the CUDA runtime, and invokes a single target kernel with independently permuted arguments (grid/block dimensions, buffer sizes, and scalar parameters) read from fuzzer-generated input files. For fair comparison, we ran these harnesses using the same \pname{} capabilities (AFL++ mutation, host-side coverage, NVBit device-side coverage in simple-trace sanitization mode) for 24 hours, repeated three times.

\cref{tab:kernel-fuzzing-comparison} summarizes the results for the 24 open-source bugs from \cref{tab:cufuzz-bugs}. Of these, kernel-level fuzzing found only six bugs (\cmark), missed eight device-side bugs (\xmark), and could not detect the ten host-side bugs by design (N/A). Focusing on the 14 device-side bugs where kernel-level fuzzing is applicable, it achieves only 43\% recall (6/14) compared to \pname{}'s whole-program approach which found all 14.

\begin{table}[!ht]
    \centering
    \caption{\rev{Kernel-level fuzzing results. \cmark{} = found, \xmark{} = missed (device-side), N/A = not applicable (host-side).}}
    \label{tab:kernel-fuzzing-comparison}
    \resizebox{0.98\textwidth}{!}{
    \begin{tabular}{llllll}
        \hline
        \rev{Bug\#} & \rev{Target} & \rev{Bug type} & \rev{Kernel/File} & \rev{Found?} & \rev{Reason} \\
        \hline
        \rev{1} & \rev{attention} & \rev{Heap buffer overflow} & \rev{reference.h:13} & \rev{N/A} & \rev{Host-side bug} \\
        \rev{2} & \rev{boxfilter} & \rev{Invalid global read} & \rev{col\_kernel} & \rev{\cmark} & \rev{Missing kernel-internal bounds check} \\
        \rev{3} & \rev{boxfilter} & \rev{Uninitialized global read} & \rev{row\_kernel} & \rev{\xmark} & \rev{Harness always initializes buffers} \\
        \rev{4} & \rev{boxfilter} & \rev{Uninitialized global read} & \rev{col\_kernel} & \rev{\cmark} & \rev{Found by malformed input parameters} \\
        \rev{5} & \rev{boxfilter} & \rev{Heap buffer overflow} & \rev{reference.cpp:121} & \rev{N/A} & \rev{Host-side bug} \\
        \rev{6} & \rev{crs} & \rev{Floating-point exception} & \rev{main.cu:81} & \rev{N/A} & \rev{Host-side bug} \\
        \rev{7} & \rev{dxtc2} & \rev{Invalid global read} & \rev{compress} & \rev{\xmark} & \rev{Host-device interface mismatch} \\
        \rev{8} & \rev{dxtc2} & \rev{Invalid global write} & \rev{compress} & \rev{\xmark} & \rev{Host-device interface mismatch} \\
        \rev{9} & \rev{dxtc2} & \rev{Data race (shared memory)} & \rev{compress} & \rev{\cmark} & \rev{Structural synchronization bug} \\
        \rev{10} & \rev{dxtc2} & \rev{Heap buffer overflow} & \rev{shrUtils.cu:1122} & \rev{N/A} & \rev{Host-side bug} \\
        \rev{11} & \rev{lud} & \rev{Invalid global read} & \rev{lud\_diagonal} & \rev{\cmark} & \rev{Found with a different trigger} \\
        \rev{12} & \rev{lud} & \rev{Stack buffer overflow} & \rev{common.cu:161} & \rev{N/A} & \rev{Host-side bug} \\
        \rev{13} & \rev{medianfilter} & \rev{Invalid global write} & \rev{ckMedian} & \rev{\cmark} & \rev{Missing kernel-internal bounds check} \\
        \rev{14} & \rev{medianfilter} & \rev{Uninitialized global read} & \rev{ckMedian} & \rev{\xmark} & \rev{Harness correctly provisions buffers} \\
        \rev{15} & \rev{medianfilter} & \rev{Data race (shared memory)} & \rev{ckMedian} & \rev{\cmark} & \rev{Harness violates pitch invariant} \\
        \rev{16} & \rev{medianfilter} & \rev{Heap buffer overflow} & \rev{MedianFilterHost.cu:70} & \rev{N/A} & \rev{Host-side bug} \\
        \rev{17} & \rev{medianfilter} & \rev{Heap buffer overflow} & \rev{MedianFilterHost.cu:85} & \rev{N/A} & \rev{Host-side bug} \\
        \rev{18} & \rev{recursiveGaussian} & \rev{Heap buffer overflow} & \rev{shrUtils.cu:1302} & \rev{N/A} & \rev{Host-side bug} \\
        \rev{19} & \rev{recursiveGaussian} & \rev{Invalid global read} & \rev{RecursiveRGBA} & \rev{\xmark} & \rev{Malformed PPM leading to dimension mismatch} \\
        \rev{20} & \rev{recursiveGaussian} & \rev{Uninitialized global read} & \rev{RecursiveRGBA} & \rev{\xmark} & \rev{Malformed PPM leading to dimension mismatch} \\
        \rev{21} & \rev{recursiveGaussian} & \rev{Uninitialized global read} & \rev{Transpose} & \rev{\xmark} & \rev{Malformed PPM leading to dimension mismatch} \\
        \rev{22} & \rev{recursiveGaussian} & \rev{Heap buffer overflow} & \rev{RecursiveGaussianHost.cu:180} & \rev{N/A} & \rev{Host-side bug} \\
        \rev{23} & \rev{seam-carving} & \rev{Invalid global read} & \rev{compute\_costs\_kernel} & \rev{\xmark} & \rev{Host-side error handling flaw} \\
        \rev{24} & \rev{urng} & \rev{Heap buffer overflow} & \rev{SDKBitMap.h:368} & \rev{N/A} & \rev{Host-side bug} \\
        \hline
    \end{tabular}
    }
    \Description{\rev{Comparison of kernel-level fuzzing results against cuFuzz findings on open-source benchmarks.}}
\end{table}

More critically, kernel-level fuzzing generated 16 \emph{false positives}, i.e., spurious errors that cannot occur in the original program (\cref{tab:kernel-fuzzing-fps}). These false alarms arise because independent parameter permutation violates host-enforced invariants. For example, in \texttt{seam-carving}'s \texttt{compute\_M\_step1} kernel, harnesses fuzz \texttt{gridSize} independently when the original program derives it from data dimensions (i.e., \texttt{gridSize = $\lceil$current\_w/128$\rceil$}). Similarly, in \texttt{lud}'s \texttt{lud\_perimeter} kernel, compile-time constants like \texttt{BLOCK\_SIZE = 16} become fuzzed runtime variables, causing shared memory sizing mismatches. These violations trigger memory errors and race conditions that are impossible under valid host-side orchestration.

\begin{table}[!ht]
    \centering
    \caption{\rev{False positives from kernel-level fuzzing. All stem from violating host-enforced parameter invariants.}}
    \label{tab:kernel-fuzzing-fps}
    \resizebox{0.95\textwidth}{!}{
    \begin{tabular}{lllll}
        \hline
        \rev{\#} & \rev{Target} & \rev{Kernel} & \rev{Error type} & \rev{Violated invariant} \\
        \hline
        \rev{FP1} & \rev{attention} & \rev{kernel1\_blockReduce} & \rev{Invalid global read} & \rev{\texttt{gridSize = n}} \\
        \rev{FP2} & \rev{attention} & \rev{kernel2\_blockReduce} & \rev{Invalid global read} & \rev{\texttt{gridSize = d}} \\
        \rev{FP3} & \rev{attention} & \rev{kernel2\_blockReduce} & \rev{Invalid global write} & \rev{\texttt{gridSize = d}} \\
        \rev{FP4} & \rev{crs} & \rev{gcrs\_m\_1\_w\_4\_coding\_dotprod} & \rev{Invalid global read} & \rev{\texttt{index} derived from buffer bounds} \\
        \rev{FP5} & \rev{crs} & \rev{gcrs\_m\_2\_w\_4\_coding\_dotprod} & \rev{Invalid global read} & \rev{\texttt{index} derived from buffer bounds} \\
        \rev{FP6} & \rev{boxfilter} & \rev{row\_kernel} & \rev{Invalid shared read} & \rev{\texttt{iRadiusAligned $\geq$ iRadius}} \\
        \rev{FP7} & \rev{boxfilter} & \rev{row\_kernel} & \rev{Invalid global read} & \rev{\texttt{gridY = uiHeight}} \\
        \rev{FP8} & \rev{urng} & \rev{noise\_uniform} & \rev{Invalid global read} & \rev{\texttt{gridSize = imageSize / blockSize}} \\
        \rev{FP9} & \rev{urng} & \rev{noise\_uniform} & \rev{Uninitialized global read} & \rev{\texttt{gridSize = imageSize / blockSize}} \\
        \rev{FP10} & \rev{lud} & \rev{lud\_perimeter} & \rev{Race condition} & \rev{\texttt{BLOCK\_SIZE = 16} (compile-time)} \\
        \rev{FP11} & \rev{lud} & \rev{lud\_perimeter} & \rev{Invalid global read} & \rev{\texttt{gridSize} derived from \texttt{matrix\_dim}} \\
        \rev{FP12} & \rev{lud} & \rev{lud\_internal} & \rev{Race condition} & \rev{\texttt{BLOCK\_SIZE = 16} (compile-time)} \\
        \rev{FP13} & \rev{lud} & \rev{lud\_internal} & \rev{Invalid global read} & \rev{\texttt{gridSize} derived from \texttt{matrix\_dim}} \\
        \rev{FP14} & \rev{seam-carving} & \rev{compute\_M\_step1} & \rev{Invalid global read} & \rev{\texttt{base\_row < h}} \\
        \rev{FP15} & \rev{seam-carving} & \rev{min\_reduce} & \rev{Invalid global write} & \rev{\texttt{gridSize} derived from output size} \\
        \rev{FP16} & \rev{seam-carving} & \rev{compute\_M\_step1} & \rev{Race condition} & \rev{\texttt{gridSize = $\lceil$current\_w/128$\rceil$}} \\
        \hline
    \end{tabular}
    }
    \Description{\rev{False positives from kernel-level fuzzing caused by independent parameter permutation.}}
\end{table}

The eight false negatives (\xmark) stem from fundamental limitations of kernel isolation. Harnesses abstract away host-side logic that is often the \emph{root cause} of device-side bugs: (1)~\textbf{file parsing failures}, where malformed images cause garbage dimensions, but harnesses always initialize buffers correctly; (2)~\textbf{error handling flaws}, where programs continue after failed \texttt{cudaMalloc} calls, but harnesses explicitly check return codes; (3)~\textbf{host-device interface mismatches}, where buggy host-side buffer sizing calculations cannot be expressed when harnesses derive buffer sizes from kernel parameters. Bug \#23 (\cref{fig:bug-23-seam-carving}) exemplifies this: the kernel error is triggered by a host-side allocation failure that kernel-level fuzzing could not reproduce.

We additionally ran kernel-level harnesses with the all-trace sanitization strategy (running all sanitizers on every input) but observed no new findings beyond those reported above, confirming that the false negative gap stems from harness design limitations rather than sanitization coverage.}

\begin{takeawaybox}
\textit{\rev{\textbf{Answer to RQ7:} Kernel-level fuzzing found only 6 of 14 device-side bugs (43\% recall) while generating 16 false positives due to violated parameter invariants. In contrast, \pname{}'s whole-program approach found all 14 device-side bugs plus 10 additional host-side bugs with zero false positives, demonstrating the importance of preserving host-device context.}}
\end{takeawaybox}

\section{Discussion} \label{sec:discussion}

\subsection{Differences Between CPU and GPU Fuzzing} 
Our experience with \pname{} reveals multiple key differences between CPU and GPU fuzzing. Most importantly, throughput differs significantly between CPU and GPU workloads. While CPU fuzzing achieves tens to hundreds of exec/sec on a single thread, GPU fuzzing is typically limited to tens of exec/sec due to CUDA runtime initialization overhead. Second, GPU fuzzing is more susceptible to context switching due to limited GPU nodes per system (8 GPUs versus 128 cores per server). Both factors limit the number and duration of fuzzing campaigns. 

Potential solutions for improving GPU fuzzing throughput include using techniques to co-execute multiple GPU processes on the same device (e.g., NVIDIA Multi-Process Service, MPS~\cite{Nvidia2025:MPS}) or radically modifying AFL++ and the application under test to process inputs in batches rather than individually. Both solutions depend on the GPU utilization of the program-under-test.  

\subsection{GPU Utilization During Fuzzing} 

Our experiments reveal that GPU utilization during fuzzing depends heavily on both the program-under-test and input characteristics. Some programs invoke multiple kernels with high thread counts, fully utilizing GPU resources, while others use only a fraction of available SMs and shared memory. Similarly, some inputs trigger deep functionality while others fail early, causing utilization to vary widely for the same program. Future work could leverage these utilization patterns to co-execute low-utilization workloads and improve throughput. 

\rev{
\subsection{Thread-Interleaving Coverage Limitations}
Our current coverage mechanism collapses thread variations---two executions with identical control flow but different thread interleavings produce the same coverage signature. Consequently, the races \pname{} finds are ``surface-level,'' exposed simply by reaching the code location where \texttt{racecheck} detects the conflict. We verified that all 5 \texttt{racecheck}-detected bugs in our evaluation were reproducible across multiple runs, suggesting they are relatively deterministic races triggered by the code structure rather than timing. However,~\pname{} may miss race conditions dependent on precise thread-scheduling timing. Developing thread-interleaving-aware coverage is a promising future direction, though naive implementation would lead to state space explosion on GPUs where thousands of threads execute concurrently. We leave this extension to future work.
}

\section{Related Work} \label{sec:related-work}
Fuzzing has long been a cornerstone of software reliability and security testing. Tools like AFL++~\cite{Fioraldi2020:AFLplusplus} and LibFuzzer~\cite{libfuzzer} demonstrate the effectiveness of coverage-guided input mutation for CPU programs. However, GPU fuzzing remains relatively underexplored due to architectural and tooling challenges unique to heterogeneous systems. Next, we review related work in GPU fuzzing, dynamic sanitizers, and other fuzzing directions.

\begin{table*}[!ht]
    \centering
    \caption{\rev{Comparison of GPU fuzzing approaches.}}
    \label{tab:related-work-comparison}
    \resizebox{0.98\textwidth}{!}{
    \begin{tabular}{l|l|l|l|l|c|c}
    \hline
    \multicolumn{1}{c|}{\rev{\multirow{2}{*}{\textbf{Tool}}}} & \multicolumn{1}{c|}{\rev{\multirow{2}{*}{\textbf{Target}}}} & \multicolumn{1}{c|}{\rev{\textbf{Execution}}} & \multicolumn{1}{c|}{\rev{\multirow{2}{*}{\textbf{Instrumentation}}}} & \multicolumn{1}{c|}{\rev{\multirow{2}{*}{\textbf{Feedback}}}} & \multicolumn{1}{c|}{\rev{\multirow{2}{*}{\textbf{Sanitizers}}}} & \rev{\textbf{Closed-source}} \\
    \multicolumn{1}{c|}{} & \multicolumn{1}{c|}{} & \multicolumn{1}{c|}{\rev{\textbf{mode}}} & \multicolumn{1}{c|}{} & \multicolumn{1}{c|}{} & \multicolumn{1}{c|}{} & \rev{\textbf{support}} \\
    \hline     
    \rev{CLFuzz~\cite{Peng2020:OpenCLFuzz}} & \rev{OpenCL kernels} & \rev{Physical GPU} & \rev{Source/IR instrumentation} & \rev{Branch (kernel)} & \rev{\xmark} & \rev{\xmark} \\
    \rev{CUDA-emulation~\cite{Singh2025:CuFuzzGT}} & \rev{CUDA kernels} & \rev{CPU emulation} & \rev{Source/IR translation} & \rev{Edge (CPU Mapped)} & \rev{\cmark~(CPU ASan)} & \rev{\xmark} \\
    \rev{Fuzz4Cuda~\cite{Xu2025:Fuzz4Cuda}} & \rev{CUDA libraries} & \rev{Physical GPU} & \rev{Debugger (CUDA-GDB)} & \rev{Basic block} & \rev{\xmark~(Crash only)} & \rev{\cmark} \\
    \rev{CUDAsmith~\cite{Jiang2020:CUDAsmith}} & \rev{CUDA compilers} & \rev{CPU} & \rev{Random code generation} & \rev{None (Black-box)} & \rev{\xmark} & \rev{N/A} \\
    \rev{Moneta~\cite{Jung2025:Moneta}} & \rev{GPU drivers} & \rev{Simulated} & \rev{Driver state recording} & \rev{State/API} & \rev{\xmark} & \rev{\cmark} \\
    \rev{DL Fuzzers~\cite{Guo2021:Audee,Wei2022:FreeFuzz,Xie2022:DocTer}} & \rev{DL frameworks} & \rev{API level} & \rev{Differential testing} & \rev{API/model coverage} & \rev{\xmark} & \rev{\cmark} \\
    \hline 
    \rev{\multirow{2}{*}{\textbf{\pname{} (Ours)}}} & \rev{CUDA programs} & \rev{\multirow{2}{*}{Physical GPU}} & \rev{Compile instrumentation (host) \& } & \rev{\multirow{2}{*}{Edge (host+device)}} & \rev{\cmark~(CPU ASan and} & \rev{\multirow{2}{*}{\cmark}} 
    \\
     & \rev{and libraries} &   & \rev{runtime instrumentation (device)}  &   & \rev{Compute Sanitizer)}  &   
    \\
    \hline
    \end{tabular}
    }
    \Description{Comparison of GPU fuzzing approaches.}
    \end{table*}

\subsection{GPU Fuzzing}

\rev{\cref{tab:related-work-comparison} summarizes the key differences between \pname{} and related GPU fuzzing approaches.} Early GPU fuzzing efforts focused on kernel-level input permutation. CVFuzz~\cite{Li2022:CVFuzz} targets OpenCL kernels by generating pathological inputs to expose complexity vulnerabilities. Similarly, CLFuzz~\cite{Peng2020:OpenCLFuzz} leverages SMT solvers to generate inputs for uncovered branches in OpenCL kernels, aiming to increase kernel-level code coverage. These approaches suffer from false positives due to the lack of host-device context awareness.	

\rev{Concurrently, Singh et al. propose hardening CUDA programs by transforming them into CPU-executable code~\cite{Singh2025:CuFuzzGT}. That approach maps GPU thread hierarchies to CPU loops to leverage mature CPU fuzzing ecosystems (e.g., AFL++, ASan). However, relying on emulation introduces significant performance overheads, with reported throughputs often falling below 1 execution per second for complex kernels, and inherently fails to capture hardware-specific behaviors. In contrast, our work performs fuzzing on physical GPU hardware. This not only achieves significantly higher throughput by running kernels natively but also ensures fidelity to the actual GPU memory consistency model and warp scheduling. Moreover, by leveraging binary instrumentation, our approach supports closed-source libraries and driver interactions that are inaccessible to transformation-based methods requiring source code or intermediate representation~\cite{Singh2025:CuFuzzGT}.}

\rev{Another concurrent work, Fuzz4Cuda~\cite{Xu2025:Fuzz4Cuda}, proposes a framework that leverages the CUDA-GDB interface to fuzz GPU libraries. While it addresses initialization latency via a persistent loop, it relies on software breakpoints to track basic block coverage, which is inherently less sensitive to logic errors than the transition-aware edge coverage utilized by our approach. This coarse granularity, combined with the inability to employ runtime sanitizers due to debugger conflicts, significantly limits its effectiveness. For instance, in a month-long campaign, Fuzz4Cuda identified only 5 bugs (primarily in nvJPEG), missing silent errors like data races and uninitialized reads. In contrast, our lightweight binary instrumentation and decoupled sanitization enabled the discovery of 43 unique vulnerabilities across 14 diverse programs and libraries in under 24 hours.}

CUDAsmith~\cite{Jiang2020:CUDAsmith} introduces a fuzzing framework for CUDA compilers. While effective for compiler validation and uncovering compiler bugs, CUDAsmith does not address runtime errors in CUDA applications or library APIs. Another orthogonal line of work focuses on GPU driver fuzzing. Moneta~\cite{Jung2025:Moneta} represents the most relevant work, statefully fuzzing GPU drivers by recalling past in-vivo driver execution states at scale. In contrast, \pname{} focuses on fuzzing user-land applications. 

Finally, while~\pname{} focuses on fuzzing programs with file-based inputs, other tools target deep learning (DL) libraries/frameworks, such as TensorFlow~\cite{Abadi2016:TensorFlow} and PyTorch~\cite{Paszke2019:PyTorch}. DL library fuzzers~\cite{Guo2021:Audee,Wei2022:FreeFuzz,Xie2022:DocTer,Christou2023:IvySyn,Deng2024:FuzzGPT} primarily depend on differential testing to expose bugs in the DL kernel and API implementations running on CPUs and/or GPUs. To the best of our knowledge, none of these tools deploy heterogeneous code coverage and sanitizers.

\subsection{GPU Dynamic Sanitizers}
Besides Compute Sanitizer~\cite{Nvidia:ComputeSanitizer}, there exist other tools for catching GPU code errors at runtime. Examples include iGUARD~\cite{Aditya2021:IGUARD}, a dynamic binary instrumentation tool based on NVBit for catching global memory data races not currently covered by Compute Sanitizer's racecheck. BinFPE~\cite{Laguna2022:BinFPE} and GPU-FPX~\cite{Li2023:GPUFPX} identify floating-point exceptions in device code, both based on NVBit. 

Compiler-based tools like cuCatch~\cite{Ziad2023:cuCatch} and HiRace~\cite{Jacobson2024:HiRace} use compiler analysis and instrumentation to capture device-side memory safety errors and data races, respectively. All these tools can be easily integrated into \pname{}'s fuzzing loop without runtime compatibility concerns.

\subsection{Other Fuzzing Directions} \label{subsec:other-fuzzing-dir}
There has been a large body of work in improving file-based fuzzers. The baseline fuzzer used in this paper, AFL++~\cite{Fioraldi2020:AFLplusplus}, aggregates many research proposals built on top of the original AFL~\cite{AFL}. However, several recent proposals are not yet integrated into mainstream AFL++. For example, data-flow-aware coverage has shown promise for improving coverage-guided fuzzing~\cite{Gan2020:GreyOne, Liang2022:PATA, Kim2023:DAFL, Herrera2023:DatAFLow}. Such techniques could be incorporated into~\pname{} to enhance host- and device-side coverage beyond the default control-flow edge coverage used in this paper.

\section{Threats to Validity} \label{subsec:threats}

\subsection{Threats to External Validity} \label{subsec:threats-external}

One external validity threat concerns the benchmark programs used in our evaluation. To mitigate this threat, we used a diverse set of programs from multiple domains that accept different input formats. Another threat involves seed selection's impact on outcomes. We addressed this by selecting inputs from each benchmark's testing suite and using identical initial seeds across all configurations. Using more seeds might yield different results (i.e., better coverage or more bugs) but would ultimately require more than 24 hours to converge given CUDA fuzzing's low throughput. 

\subsection{Threats to Internal Validity} \label{subsec:threats-internal}

The main internal validity threat is randomness in the fuzzing process, which could introduce experimental result variability. We addressed this by repeating all experiments three times, using identical initial seeds across all configurations, and employing the ``-Z'' flag to force sequential exploration of the fuzzing input queue. 

\section{Conclusion} \label{sec:conclusion}

The rapid adoption of GPU computing in domains ranging from high-performance computing to safety-critical systems has amplified the importance of robust GPU testing. However, the unique characteristics of GPU architectures, such as massive parallelism, heterogeneous memory hierarchies, and complex host-device interactions, introduce distinct challenges for identifying memory- and thread-safety vulnerabilities. Fuzzing is a proven automated testing method for identifying these vulnerabilities on CPUs, yet it remains underutilized for GPUs. 

\pname{} demonstrates that practical and effective fuzzing of CUDA-based GPU programs is feasible. By rethinking fuzzing granularity, integrating device-side coverage, and decoupling sanitization from coverage collection, \pname{} overcomes multiple barriers in GPU testing. Its ability to uncover~43 previously unknown bugs---including 19 in widely deployed production libraries---highlights the inadequacy of existing testing pipelines. As GPU workloads continue to expand into privacy-sensitive applications, robust testing tools like \pname{} will be essential. 

\section*{Data-Availability Statement} \label{sec:data-availability}

The \pname{} artifact is available on Zenodo~\cite{cufuzz-artifact}. The artifact includes the source code, usage instructions, and evaluation scripts to reproduce the main experiments in this paper.

\begin{acks}
We thank the anonymous reviewers for their constructive feedback that helped improve this paper. We are grateful to Zheming Jin for promptly addressing the HeCBench bug reports uncovered by \pname{}, and to the maintainers of NVIDIA's CUDA-accelerated libraries for handling the reported bugs. We also thank Aamer Jaleel, Mark Stephenson, Sana Damani, and the members of the Architecture Research Group at NVIDIA Research for helpful technical discussions.
\end{acks}

\bibliographystyle{ACM-Reference-Format}
\bibliography{refs}

\end{document}